\documentclass[pdftex,twocolumn,epjc3,final]{svjour3}
\pdfoutput=1

\usepackage{fixltx2e}
\usepackage[normalem]{ulem}
\usepackage{placeins}
\usepackage{ragged2e}
\usepackage{url}
\usepackage[hyperindex,breaklinks]{hyperref}
\hypersetup{
        colorlinks,
                citecolor=black,
                filecolor=black,
                linkcolor=black,
                urlcolor=black
}

\usepackage{preprintcover}  
\PreprintCoverPaperTitle{Measurement of the production and lepton charge asymmetry of \textbf{\Wboson}\ bosons in \PbPb\ collisions at $\mathbf{\sqrt{\mathbf{s}_{\mathrm{\mathbf{NN}}}}=2.76\;TeV}$ with the ATLAS detector}  
\PreprintIdNumber{CERN-PH-EP-2014-156}  
\PreprintCoverAbstract{A measurement of $\Wboson$ boson production in lead-lead collisions at
\energy\ is presented. It is based on the analysis of data
collected with the ATLAS detector at the LHC in 2011 corresponding to
an integrated luminosity of 0.14 $\mathrm{nb}^{-1}$ and 0.15 $\mathrm{nb}^{-1}$ in the muon and
electron decay channels, respectively.
The differential production yields and lepton charge asymmetry are each measured as a function of the average number 
of participating nucleons $\mNpart$ and absolute pseudorapidity of the charged lepton. 
The results are compared to predictions based on next-to-leading-order QCD calculations. 
These measurements are, in principle, sensitive to possible nuclear modifications to the parton distribution functions and also provide information on scaling of $\Wboson$ boson production in multi-nucleon systems.}
\PreprintJournalName{EPJ C}  

\usepackage{graphicx}
\usepackage{atlasphysics}

\usepackage[T1]{fontenc}
\usepackage{graphicx}
\usepackage{mathptmx}     
\usepackage{flushend}
\usepackage{placeins}
\usepackage{amsmath}

\journalname{Eur. Phys. J. C}
\usepackage[normalem]{ulem}
\usepackage{color}

\newcommand{\energy}{$\sqrt{s_{\mathrm{NN}}}=2.76\tev$}
\newcommand{\PbPb}{\mbox{Pb+Pb}}
\newcommand{\pPb}{\mbox{$p$+Pb}}
\newcommand{\pp}{\mbox{$\emph{pp}$}}
\newcommand{\nn}{\mbox{$\emph{nn}$}}
\newcommand{\pn}{\mbox{$\emph{pn}$}}

\newcommand{\Npart}{\mbox{$N_{\mathrm{part}}$}}
\newcommand{\Ncoll}{\mbox{$N_{\mathrm{coll}}$}}
\newcommand{\sumet}{\mbox{$\Sigma E_{\mathrm{T}}$}}
\newcommand{\pho}{\phantom{0}}

\newcommand{\imub}{\ensuremath{\mu \mathrm{b}^{-1}}}
\newcommand{\mt}{m_{\mathrm{T}}}
\newcommand{\mex}{E_{\textrm{x}}^{\textrm{miss}}}
\newcommand{\mey}{E_{\textrm{y}}^{\textrm{miss}}}

\newcommand{\mpt}{\pt^{\mathrm{miss}}}

\newcommand{\mNpart}{\langle N_{\mathrm{part}} \rangle}
\newcommand{\mNcoll}{\langle N_{\mathrm{coll}} \rangle}
\newcommand{\taa}{\langle T_{AA} \rangle}

\newcommand{\mtEqn}{m_{\mathrm{T}} = \sqrt{2\pt^{\ell}\mpt(1-\cos\Delta\phi_{\ell,\mpt})}}
\newcommand{\wmunu}{\Wboson\rightarrow\mu\nu_{\mu}}
\newcommand{\wenu}{\Wboson\rightarrow e\nu_{e}}
\newcommand{\wlnu}{\Wboson\rightarrow\ell\nu_{\ell}}
\newcommand{\wmunup}{\Wboson^{+}\rightarrow\mu^{+}\nu_{\mu}}
\newcommand{\wmunum}{\Wboson^{-}\rightarrow\mu^{-}\bar{\nu_{\mu}}}

\newcommand{\wenup}{\Wboson^{+}\rightarrow e^{+}\nu_{e}}
\newcommand{\wenum}{\Wboson^{-}\rightarrow e^{-}\bar{\nu_{e}}}

\newcommand{\wlnup}{\Wboson^{+}\rightarrow\ell^{+}\nu_{\ell}}
\newcommand{\wlnum}{\Wboson^{-}\rightarrow\ell^{-}\bar{\nu_{\ell}}}

\newcommand{\wtaunu}{\Wboson\rightarrow\tau\nu_{\tau}}
\newcommand{\wtaumu}{\Wboson\rightarrow\tau\nu_{\tau}\rightarrow\mu\nu_{\mu}\nu_{\tau}\nu_{\tau}}
\newcommand{\wtaue}{\Wboson\rightarrow\tau\nu_{\tau}\rightarrow e\nu_{e}\nu_{\tau}\nu_{\tau}}
\newcommand{\wtaul}{\Wboson\rightarrow\tau\nu_{\tau}\rightarrow\ell\nu_{\ell}\nu_{\tau}\nu_{\tau}}
\newcommand{\ztautau}{\Zboson\rightarrow\tau\tau}
\newcommand{\zee}{\Zboson\rightarrow e^+e^- }

\newcommand{\zmumu}{\Zboson\rightarrow\mu^{+}\mu^{-}}

\newcommand{\tauola}{{\sc Tauola}}
\newcommand{\geant}{{\sc Geant4}}
\newcommand{\pythia}{{\sc Pythia}}
\newcommand{\powheg}{{\sc Powheg}}
\newcommand{\hijing}{{\sc Hijing}}

\newcommand{\photos}{{\sc Photos}}

\usepackage{txfonts}
\usepackage{multirow}
\usepackage[numbers,sort&compress]{natbib}
\begin{document}

\title{Measurement of the production and lepton charge asymmetry of \textbf{\Wboson}\ bosons in \PbPb\ collisions at $\mathbf{\sqrt{\mathbf{s}_{\mathrm{\mathbf{NN}}}}=2.76\;TeV}$ with the ATLAS detector}

\author{The ATLAS Collaboration}
\institute{The ATLAS Collaboration \at CERN, 1211 Geneva 23, Switzerland}

\date{Received: date / Accepted: date}
\titlerunning{\Wboson\ boson production and lepton charge asymmetry in \PbPb\ collisions with the ATLAS detector}
\maketitle 
\xspaceskip=0.4em
\begin{abstract} 
A measurement of $\Wboson$ boson production in lead-lead collisions at
\energy\ is presented. It is based on the analysis of data
collected with the ATLAS detector at the LHC in 2011 corresponding to
an integrated luminosity of 0.14 $\mathrm{nb}^{-1}$ and 0.15 $\mathrm{nb}^{-1}$ in the muon and
electron decay channels, respectively.
The differential production yields and lepton charge asymmetry are each measured as a function of the average number 
of participating nucleons $\mNpart$ and absolute pseudorapidity of the charged lepton. 
The results are compared to predictions based on next-to-leading-order QCD calculations. 
These measurements are, in principle, sensitive to possible nuclear modifications to the parton distribution functions and also provide information on scaling of $\Wboson$ boson production in multi-nucleon systems.
\end{abstract}

\section{Introduction}
\label{sec:intro}

Studies of particle production in the high--density medium created in
ultra--relativistic heavy--ion collisions have been previously conducted at the Relativistic Heavy Ion Collider (RHIC) at Brookhaven National Laboratory~\cite{Arsene:2004fa,Back:2004je,Adams:2005dq,Adcox:2004mh} 
and have been extended to larger centre--of--mass energies at the Large Hadron Collider~(LHC) at CERN~\cite{ATLAS:2011ag,Chatrchyan:2011pb}.
These collisions provide access to a phase of nuclear matter at high temperature and low baryon density called quark--gluon plasma (QGP), 
in which the relevant degrees of freedom are quarks and gluons~\cite{Tannenbaum:2006ch,Mohanty:2009vb,Fukushima:2010bq,Mohanty:2011ki,Muller:2012zq}.  
In a QGP, high--energy partons transfer energy to the medium through multiple interactions and gluon radiation, resulting in a modification of the parton shower of jets~(jet--quenching).
This effect is consistent with the measurements of high transverse momentum~($\pt$) charged hadron yields~\cite{Adler:2003qi,Adams:2003kv,Aamodt:2010jd,CMS:2012aa,Aad:2014wha},
inclusive jets~\cite{Aad:2012vca} and dijets with asymmetric transverse energies~($\et$)~\cite{Aad:2010bu,Chatrchyan:2011sx,Chatrchyan:2012nia}.

Electroweak bosons~($V = \gamma$, $\Wboson$, $\Zboson$) provide additional ways to study partonic energy loss in heavy--ion collisions. They do not interact strongly with the medium, thus offering 
a means to calibrate the energy of jets in $V$--jet events.
At sub--TeV centre--of--mass energies, the only viable candidates for playing this role are photons~\cite{Afanasiev:2012dg}. However at higher energies, 
heavy gauge bosons~($\Wboson^{\pm}$ and $\Zboson$) are also produced in relatively high abundance, introducing an additional avenue for
benchmarking in--medium modifications to coloured probes.
This potential has already been realised in lead--lead~($\PbPb$) collisions in previous 
ATLAS~\cite{Aad:2012ew} and CMS~\cite{Chatrchyan:2011ua,Chatrchyan:2012vq,Chatrchyan:2012nt} publications, where it was observed that 
electroweak boson production rates scale linearly with the number of binary nucleon--nucleon collisions.

Moreover, in principle, electroweak bosons are an excellent tool for studying modifications to parton distribution functions~(PDFs) in a multi--nucleon environment.
To leading--order, $W^{+}(W^{-})$ bosons are primarily produced by interactions between a $u(d)$ valence quark and a $\overline{d}(\overline{u})$ sea quark.
The rapidity of the $\Wboson$ boson is primarily determined by the momentum fractions, $x$,
 of the incoming partons. 
 Therefore, information about the PDF can be extracted by measuring the charge asymmetry as a function of the pseudorapidity\footnote{The ATLAS detector uses a right--handed coordinate system with the nominal $\PbPb$ interaction point at its centre.
The $z$--axis is along the beam pipe. The $x$--axis points from the interaction point toward the centre of the ring and the $y$--axis points upward. Cylindrical 
coordinates ($r$,~$\phi$) are used in the transverse plane with $\phi$ being the azimuthal angle around the $z$--axis. The pseudorapidity is defined in terms of the polar angle $\theta$ as $\eta=-\ln(\tan{\theta}/2)$.} 
of charged leptons produced from $\Wboson$ decays.

The charge asymmetry is defined in terms of the differential production yields for $\wlnu$ ($\ell=\mu, e$), $\mathrm{d}N_{\wlnu}/\mathrm{d}\eta_{\ell}$: 

\begin{equation}
 A_{\ell}(\eta_{\ell})= \frac{\mathrm{d}N_{\wlnup}/\mathrm{d}\eta_{\ell}-\mathrm{d}N_{\wlnum}/\mathrm{d}\eta_{\ell}}{\mathrm{d}N_{\wlnup}/\mathrm{d}\eta_{\ell}+\mathrm{d}N_{\wlnum}/\mathrm{d}\eta_{\ell}}
\label{eqasym}
\end{equation}

\noindent
where $\eta_{\ell}$ is the pseudorapidity
of the charged lepton and the $\Wboson$ boson production yields are determined in the kinematic phase space used to select $\wlnu$ events.
This observable has been used to study PDFs in binary nucleon systems such as \pp\ collisions at the LHC~\cite{Aad:2011dm,Chatrchyan:2012xt,Chatrchyan:2013mza} and $p\bar{p}$ collisions at the Tevatron~\cite{Aaltonen:2009ta,Abazov:2013rja}.  
However, its utility in nuclear systems has only recently been explored with a limited set of experimental data~\cite{Chatrchyan:2012nt}.

Although the method for measuring the charge asymmetry in $\PbPb$
is essentially identical to that in \pp,  the distributions themselves are not expected to be identical. 
In \pp\ collisions, the overall production rate of $W^{+}$ bosons is larger than that of $W^{-}$ bosons as a result of the  
larger fraction of $u$ valence quarks relative to $d$ valence quarks in the colliding system. 
On the other hand, in \PbPb\ collisions, the nuclei contain 126 neutrons and 82 protons. Thus, \pp\ interactions make up 
only $\approx$~15\% of the total number of nucleon--nucleon interactions, whereas neutron--neutron (\nn) and proton--neutron (\pn) combinations contribute $\approx$~37\% and $\approx$~48\%, respectively.
Consequently, a marked difference is expected in the lepton charge asymmetry between \PbPb\ and \pp\ collisions.

Prior to this analysis, the only published charge asymmetry measurement in heavy--ion collisions was reported by the CMS collaboration~\cite{Chatrchyan:2012nt} with an integrated luminosity of $7.3~\imub$ using the $\wmunu$ channel in \PbPb\ collisions at \energy. 
The measurement presented here uses a dataset from 2011, which corresponds to an integrated luminosity of 0.14 and 0.15$~\inb$ for the muon and electron channels, respectively. 
In addition, the $\wenu$ decay mode is employed for the first time in a heavy--ion environment. 

The paper is organised as follows: 
a brief overview of the ATLAS detector and trigger is given in Sect.~\ref{sec:detector}. 
A description of the simulated event samples used in the analysis is provided in Sect.~\ref{sec:mc}. 
The criteria for selecting $\PbPb$ events are presented in Sect.~\ref{sec:event-sel}. 
This is followed by a description of muon and electron reconstruction and signal candidate selection in Sect.~\ref{sec:reco}. 
The background estimations are presented in Sect.~\ref{sec:bkg}. 
A discussion of the procedure for correcting the signal yields is presented in Sect.~\ref{sec:correction}. The systematic uncertainties and the combination of the two channels are described in Sect.~\ref{sec:syst}, and  
the $\Wboson$ boson production yields, measured as a function of the mean number of inelastically interacting nucleons $\mNpart$ and $|\eta_\ell|$, are discussed in Sect.~\ref{sec:results}. 
A differential measurement of the lepton charge asymmetry as a function of $|\eta_\ell|$ is also presented. 
These results are compared to predictions at next--to--leading order~(NLO)~\cite{Altarelli:1979ub,KubarAndre:1978uy,Kubar:1980zv} in QCD, both with and without nuclear corrections. The former is represented by the EPS09 PDF~\cite{Paukkunen:2010qg}. 
Section~\ref{sec:summary} provides a brief summary of the results. 

\section{The ATLAS detector}
\label{sec:detector}

ATLAS~\cite{Aad:2008zzm}, one of four large LHC experiments,  is well equipped to carry out an extensive heavy--ion program. The inner detector~(ID) comprises a precision tracking system that covers a pseudorapidity range $|\eta|<2.5$. 
The ID consists of silicon pixels, silicon microstrips, and a transition radiation tracker~(TRT)\footnote{The TRT provides tracking information up to $|\eta|<2$.} consisting of cylindrical drift tubes and operates within a 2~T axial magnetic field supplied by a superconducting solenoid. 

Due to the high occupancy in heavy--ion events, tracks of charged particles are reconstructed using only the silicon pixels and microstrips. No information from the TRT is used in this analysis, and henceforth ID tracks will refer to those tracks that are reconstructed without this detector component.  

Outside the solenoid, highly segmented electromagnetic~(EM) and hadronic sampling calorimeters cover the region $|\eta|<4.9$.
The EM calorimetry is based on liquid--argon (LAr) technology and is divided into one barrel ($|\eta|<1.475$, EMB) and two end--cap ($1.375<|\eta|<3.2$, EMEC) components. The transition region between the barrel and end--cap calorimeters is located within the pseudorapidity range $1.37<|\eta|<1.52$. 
The hadronic calorimeter is based on two different
detector technologies: steel absorber interleaved with plastic scintillator covering the barrel ($|\eta|<1.0$) and extended barrels ($0.8<|\eta|<1.7$) 
and LAr hadronic end--cap calorimeters (HEC) located in the region $1.5<|\eta|<3.2$.  
A forward calorimeter (FCal) that uses LAr as the active material is located in the region $3.1<|\eta|<4.9$.  
On the inner face of the end--cap calorimeter cryostats, a minimum-bias trigger scintillator~(MBTS) is installed on each side of the ATLAS detector, covering the pseudorapidity region $2.1<|\eta|<3.8$.    

The outermost sub-system of the detector is a muon spectrometer~(MS) that is divided into a barrel region ($|\eta|<1.05$) and two end--cap regions ($1.05<|\eta|<2.7$). 
Precision measurements of the track coordinates and momenta are provided by monitored drift tubes~(MDTs), 
cathode strip chambers (CSCs), and 
three sets of air--core superconducting toroids with coils arranged in an eight--fold symmetry that provide on average 0.5 T in the azimuthal plane.  

The zero--degree calorimeters~(ZDCs)~\cite{Jenni:1009649} are located symmetrically at $z=\pm140$~m and cover $|\eta|>8.3$. In \PbPb\ collisions the ZDCs primarily measure spectator neutrons from the colliding nuclei.

The ATLAS detector also includes a three--level trigger system~\cite{Aad:2012xs} : level one~(L1) and the software--based High Level Trigger (HLT), which is subdivided into the Level 2 (L2) trigger and Event Filter (EF).
Muon and electron triggers are used to acquire the data analysed in this paper. 

The trigger selection for muons is performed in three steps. 
Information is provided to the L1 trigger system by the fast--response resistive plate chambers~(RPCs) in the barrel~($|\eta|<1.05$) and thin gap chambers~(TGCs) in the end--caps~($1.05<|\eta|<2.4$). Both the RPCs and TGCs are part of the MS.
Information from L1 is then passed to the HLT, which  
reconstructs muon tracks  
in the vicinity of the detector region reported by the L1 trigger. The L2 trigger performs a fast reconstruction of muons 
using a simple algorithm, which is then further refined at the EF by utilising the full detector information as in the offline muon reconstruction software. 

The trigger selection for electrons is performed using a L1 decision based on electromagnetic energy depositions in trigger towers of $\Delta \phi \times\Delta\eta = 0.1\times 0.1$ formed by EM calorimeter cells within the range $|\eta|<2.5$.
The electron trigger algorithm identifies a region of interest as a trigger tower cluster for which the transverse energy~(\et) sum from at least one of the four possible pairs of nearest neighbour towers exceeds a specified $\et$ threshold.

\section{Monte Carlo samples}
\label{sec:mc}

Simulated event samples are produced using the Monte Carlo~(MC) method and are used to estimate both the signal and background components. 
The response of the ATLAS detector is simulated using \geant~\cite{Agostinelli:2002hh,Aad:2010ah}.
The samples used throughout this paper are summarised in Table~\ref{tab:mcsamples}.
Each signal process and most of the background processes are embedded into minimum--bias~(MB) heavy--ion events from data recorded in the same run periods as the data used to analyse $\Wboson$ boson production. Events from the $\zmumu$ channel are embedded into \hijing~\cite{Wang:1991hta} -- a widely used heavy--ion simulation that reproduces many features of the underlying event~\cite{Aad:2012vca}.

\begin{table}[!htb]\caption{Signal and background simulated event samples used in this analysis. $\wlnu$ events include all nucleon combinations, whereas background processes use only \pp\ simulations. The variable $\hat{p}_{\mathrm{T}}$ is 
the average $\pt$ of the two outgoing partons involved in the hard--scattering process evaluated before modifications from initial-- and final--state radiation. Details for each sample are given in the text.}
\label{tab:mcsamples}
\begin{center}
\begin{tabular}{l|l|l}
\hline
Physics process & Generator& PDF set \\
\hline
$\wmunu$ & \powheg+\pythia8 & CT10 \\
$\wenu$ & \powheg+\pythia8 & CT10 \\
Dijet & \pythia6 & MRST LO* \\
\hspace*{0.05cm} ($17<\hat{p}_{\mathrm{T}}<140~\GeV$) & & \\

$\zmumu$ & \pythia6 & MRST LO* \\
$\zee$ & \powheg+\pythia8 & CT10 \\
$\wtaumu$ & \pythia6 & MRST LO* \\
$\wtaue$ & \powheg+\pythia8 & CT10 \\
\hline
\end{tabular}
\end{center}
\end{table}

The production of $\Wboson$ bosons and its decay products are modelled with the \powheg~\cite{Alioli:2008gx} event generator, which is interfaced to \pythia8~\cite{Sjostrand:2007gs} in order to model parton showering and fragmentation processes. These samples use the CT10~\cite{Lai:2010vv} PDF set and are used to estimate the signal selection efficiency and to provide predictions from theory. 
In order to account for the isospin of the nucleons, separate samples of \pp, \pn, and \nn\ events are generated and combined in proportion to their corresponding collision frequency in \PbPb\ collisions.
Only \pp\ simulations are used to model background processes (discussed in detail in Sect.~\ref{sec:bkg}) since these channels are not sensitive to isospin effects.  

Background samples are generated for muons with \pythia6 using the MRST LO* PDF set~\cite{Sherstnev:2007nd} and for electrons with \powheg\ using the CT10 PDF set.
At the level of the precision of the background estimation, no significant difference is expected between the \pythia6\ and \powheg\ generators. 
The background contribution to the muon channel from heavy--flavour is modelled using simulated dijet samples 
with average final-state parton energies $\hat{p}_{\mathrm{T}}$ in the range 17--140~\GeV.
Tau decays from $\wtaunu$ events are treated using either \tauola~\cite{Jadach:1990mz} 
or \pythia8 for final states involving muons or electrons, respectively. 
Final--state radiation from QED processes is simulated by \photos~\cite{Golonka:2005pn}.

\section{Event selection}
\label{sec:event-sel}
\subsection{Centrality definition} 
\PbPb\ collision events are selected by imposing basic requirements on the beam conditions and the performance of each sub--detector.
In order to select MB hadronic \PbPb\ collisions, a hit on each side of the MBTS system with a time coincidence within 3~ns is required for each collision.
In addition, each event is required to have a reconstructed vertex with at least three associated high--quality tracks~\cite{ATLAS:2011ah} compatible with the beam--spot position. 
These requirements select MB hadronic \PbPb\ collisions in the data with an efficiency of $(98\pm2)\%$ with respect to the total non--Coulombic inelastic cross--section~\cite{ATLAS:2011ag}. After accounting for the selection efficiency and prescale factors imposed by the trigger system during data taking~\cite{ATLAS-CONF-2012-122}, approximately $1.03 \times10^{9}$ \PbPb\ events are sampled~(denoted by $N_{\mathrm{events}}$ hereafter).

Each event is categorised into a specific centrality class defined by selections on FCal \sumet, the
total transverse energy deposited in the FCal and calibrated to the EM energy scale~\cite{ATLAS:2011ah}. 
Centrality classes in heavy--ion events represent the percentiles of the total inelastic non--Coulombic \PbPb\ cross--section. 
This reflects the overlap volume between the colliding nuclei and allows for selection of various collision geometries in the initial state. 

The FCal \sumet\ is closely related to the mean number of inelastically interacting nucleons $\mNpart$ and mean number of binary collisions $\mNcoll$ through the Glauber formalism~\cite{Miller:2007ri}. 
$\mNpart$ and $\mNcoll$ are monotonic functions of the collision impact parameter and are correlated with the FCal \sumet\ of each \PbPb\ collision~\cite{ATLAS:2011ag}. 
$\mNcoll$ can also be expressed as the product of the average nuclear thickness function $\taa$ and the total inelastic \pp\ cross--section ($64\pm5$ mb at $\sqrt{s}=2.76\tev$~\cite{pdg_2013}). In this paper, events are separated into five centrality classes: 0--5\%, 5--10\%, 10--20\%, 20--40\%, and 40--80\% with the most
central interval~(0--5\%) corresponding to the 5\% of events with the largest
FCal \sumet. 
The $\mNcoll$ estimation in the 80--100\% class suffers from high experimental uncertainties, and therefore, this centrality class is not considered in the analysis.
Table~\ref{tab:centrality} presents $\mNpart$ and $\mNcoll$ for each centrality class along with their relative systematic uncertainties~(see Sect.~\ref{sec:syst}).
Since a single participant can interact inelastically with several nucleons in a collision, the uncertainty in $\mNpart$ is less than that of the corresponding $\mNcoll$ in each centrality class. 

\begin{table}[!htb]
\caption{ Average number of participating nucleons $\mNpart$ and 
          binary collisions $\mNcoll$ for the centrality classes used in this analysis
          alongside their relative uncertainties.}
\label{tab:centrality}
\begin{center}
\begin{tabular}{c|cc|cc}
\hline
Centrality [\%] & $\langle\Npart\rangle$ & $\delta\langle\Npart\rangle$ [\%] & $\langle\Ncoll\rangle$ & $\delta\langle\Ncoll\rangle$ [\%]\\
\hline
0--5 &382 &0.5 &1683 & 7.7  \\
5--10 & 330 & 0.9&1318 & 7.5       \\
10--20 & 261 & 1.4& 923 & 7.4      \\
20--40 &158 & 2.6& 441 & 7.3       \\
40--80 & 46 & 6.0 & 78 & 9.4       \\
\hline
0--80 & 140 & 4.7 & 452 & 8.5 \\

\hline
\end{tabular}
\end{center}
\end{table}

\subsection{Trigger selection} 
$\wmunu$ candidates are selected using single muon triggers with a requirement on the minimum transverse momentum of $10 \GeV$ in the HLT.
Two types of single muon triggers are used: one that requires a muon in coincidence with a total event transverse energy  -- measured in the calorimeter at L1 -- above  $10 \GeV$ and another which requires a muon in coincidence with a neutral particle at $|\eta|>8.3$ in the ZDCs. This combination of triggers maximises the efficiency for events across all centrality classes. 
The muon trigger efficiencies are evaluated using high--quality single muons reconstructed from MB events and range from 89.3\% to 99.6\%, 
depending on $|\eta_{\mu}|$ and the centrality of the event from which the muon originated.

Candidate events for $\wenu$ are selected using only the hardware--based L1 trigger, i.e. without use of the HLT. 
The L1 calorimeter trigger selects photon and electron candidates in events where the transverse energy in an EM cluster of trigger towers exceeds 14~GeV. 
The efficiency is evaluated using a tag--and--probe method that utilises $\zee$ events selected using the criteria from Ref.~\cite{Aad:2012ew}. 
This gives an efficiency of 99.6\% for electrons with $\et>25\gev$ and $|\eta|<2.47$ -- excluding the transition region -- with a negligible centrality dependence. 

\subsection{Transverse momentum imbalance, $\mpt$} 
Previous $\Wboson$ boson analyses in
ATLAS~\cite{Aad:2011dm} have used the event momentum imbalance in the plane transverse to the beam axis~($\met$) as a proxy for 
the true neutrino $\pt$. Traditionally, these analyses reconstruct the $\met$ using contributions from energy 
deposits in the calorimeters and muons reconstructed in the MS~\cite{ATLAS-CONF-2012-101}. 
In minimum bias events, no genuine missing energy is expected, and the resolution of the two $\met$ components ($\sigma^{\mathrm{miss}}_{x}$, $\sigma^{\mathrm{miss}}_{y}$)
is measured directly from reconstructed quantities in the data by assuming the true $\mex$ and $\mey$ are zero. The resolution 
is estimated from the width of the $\mex$ and $\mey$ distributions. 
In heavy--ion collisions, 
soft particle production is much higher than in \pp\ collisions, thereby resulting in an increased number of particles 
that do not reach the calorimeter or seed a topocluster. 
Consequently,  
the resolution in the $\met$ observed in the data using calorimeter cells is at the level of $45\gev$ in the most central heavy--ion events. 
Therefore, this analysis employs a track--based calculation proposed in Ref.~\cite{Chatrchyan:2012nt} that provides a four--fold improvement in resolution relative to the calorimeter--based method.
The event momentum imbalance using this approach is 
defined as the negative vector sum of all high--quality ID tracks~\cite{ATLAS:2011ah} with $\pt>3 \GeV$:

\begin{equation}
\mathbf{p}^{\mathrm{miss}} = -\sum_{\mathrm{i}=1}^{N_\mathrm{{tracks}}}{\displaystyle \mathbf{p}^{\mathrm{track}}_\mathrm{i}},
\label{eqnMPT}
\end{equation}
\noindent where $\mathbf{p}^{\mathrm{track}}_\mathrm{i}$ is the momentum vector of the $i^{\mathrm{th}}$ ID track, and $N_\mathrm{{tracks}}$ represents the total number of ID tracks in the event.
The magnitude of the transverse component $\mpt$ and azimuthal angle $\phi^{\mathrm{miss}}$ are calculated from the transverse components ($p_{\mathrm{x}}^{\mathrm{miss}}$ and $p_{\mathrm{y}}^{\mathrm{miss}}$) of the resultant vector.
The lower track $\pt$ threshold is chosen based on that which gives the best resolution in the $\mpt$ while still including a sufficient number of tracks in the vector summation.

The transverse mass of the charged lepton and neutrino system is defined as
\begin{equation}
 \mtEqn,
\end{equation}
where $\Delta\phi_{\ell,\mpt}$ is the difference between the direction of the charged lepton and $\mpt$ vector in the azimuthal plane. 

\section{Signal candidate reconstruction and selection}
\label{sec:reco}

\subsection{Muon reconstruction}
\label{sec:mu-reco}

Muon reconstruction in ATLAS consists of separate tracking in the ID and MS. 
In this analysis, tracks reconstructed in each sub--system are combined using the $\chi^{2}$-minimisation procedure described in Ref.~\cite{ATLAS-CONF-2010-036}.
These combined muons are required to satisfy selection criteria that closely follow those used in the $\Zboson$ boson analysis in \PbPb\ data~\cite{Aad:2012ew}.
To summarise, these criteria include a set of ID hit requirements in the pixel and SCT layers of the ID, 
a selection on the transverse and longitudinal impact parameters ($|d_{0}|$ and $|z_{0}|$),
and a minimum requirement on the quality of the muon track fit. 
Additional selection criteria specific to $\Wboson$ bosons are discussed below. 

Decays--in--flight from pions and kaons contribute a small background fraction in this analysis. 
They are reduced by requiring the difference between the ID and MS muon $\pt$ measurements (corrected for the mean energy loss due to 
interactions with the material between the ID and MS) to be less than 50\% of the $\pt$ measured in the ID. 
Decays--in--flight are further reduced by locating changes in the direction of the muon track trajectory. This is performed 
using a least--squares track fit that includes scattering angle parameters accounting for multiple scattering between the muon and detector material.  
Scattering centers are allocated along the muon track trajectory from the ID to MS, and decays are identified by scattering angle measurements much 
greater than the expectation value due to multiple scattering~\cite{ATLAS-CONF-2011-003}.

In order to reduce the multi--jet contribution,
a track--based isolation of the muon is imposed. 
The tracks are taken from a cone radius $\Delta R=\sqrt{(\Delta\eta)^2+(\Delta\phi)^2}=0.2$ around the direction of the muon. 
The muon is considered isolated if the sum of the transverse momenta of ID tracks~($\sum\pt^{\mathrm{ID}}$) with $\pt>3\GeV$ -- excluding the muon $\pt$ itself -- is 
less than 10\% of the muon $\pt$.
In this paper, the quantity $\sum\pt^{\mathrm{ID}}/\pt$ is referred to as the muon isolation ratio.
Based on MC studies, 
the isolation requirement is estimated to reject 50--70\% of 
muons in QCD multi--jet events, depending on the centrality class, 
while retaining at least 95\% of signal candidates.

\subsection{Electron reconstruction}
\label{sec:e-reco}

In order to reconstruct electrons in the environment of heavy--ion collisions, the energy deposits from soft particle production due to the underlying event~(UE) must be subtracted, as they distort calorimeter--based observables.
 The two--step subtraction procedure, described in detail in Ref.~\cite{Aad:2012vca}, is applied. It involves calculating a per--event average UE energy density that excludes contributions from jets and EM clusters and accounts for effects from elliptic flow modulation on the UE. 
 The residual deposited energies stem primarily from three sources: photons/electrons, jets and UE fluctuations~(including higher--order flow harmonics). After the UE background subtraction, a standard ATLAS electron reconstruction and identification algorithm~\cite{Aad:2011mk,Aad:2014fxa} for heavy--ions is used -- the only difference between this algorithm and the one used in $\pp$ collisions is that the TRT is not used. The algorithm is designed to provide various levels of background rejection and high identification efficiencies over the full acceptance of the ID system.

 The electron identification selections are based on criteria that use calorimeter and tracking information and are optimised in bins of $\eta$ and \et. Patterns of energy deposits in the first layer of the EM calorimeter, track quality variables, and a cluster--track matching criterion are used to select electrons.  
Selection criteria based on shower shape information from the second layer of the EM calorimeter and energy leakage into the hadronic calorimeters are used as well.
Background from charged hadrons and secondary electrons from conversions are reduced by imposing a requirement on the ratio of cluster energy to track momentum. Electrons from conversions are further reduced by requiring at least one hit in the first layer of the pixel detector.

A calorimeter--based isolation variable is also imposed. Calorimeter clusters are taken within $\Delta R = 0.25$ around the candidate electron cluster. An electron is considered isolated if the total transverse energy of calorimeter clusters -- excluding the candidate electron cluster -- is less than 20\% of the electron \et. In this paper, the quantity $\sum\et^{\mathrm{calo}}/\et$ is referred to as the electron isolation ratio. The isolation requirement was studied in each centrality class and retains, on average, 92\% of signal candidates while rejecting 42\% of electrons from QCD multi--jet events.

\subsection{$\Wboson$ boson candidate selection}
\label{sec:w-selection}

$\Wboson$ boson production yields are measured in a fiducial region defined by: 

\vspace{3 mm}
\begin{tabular}{@{\hspace{-1.6em}}l@{\hspace{2em}}c@{\hspace{1em}}l}
  $\wmunu$: & $\pt^{\mu}>25\GeV$, & $0.1<|\eta_{\mu}|<2.4$, \\
  \pho{} & $\pt^{\nu}>25\GeV$, & $\mt>40\GeV$; \\
  $\wenu$: & $\pt^{e}>25\GeV$, & $|\eta_{e}|<2.47$, \\
  \pho{} & \multicolumn{2}{@{\hspace{0.0em}}l}{excluding $1.37<|\eta_{e}|<1.52$,} \\ 
  \pho{} & $\pt^{\nu}>25\GeV$, & $\mt>40\GeV$. \\
\end{tabular}
\vspace{3 mm}
\\*

In the MS, a gap in chamber coverage is located at $|\eta_{\mu}|<0.1$ that allows for services to the solenoid magnet, calorimeters, and ID, and therefore, this region is excluded. 
The most forward bin boundary is determined by the acceptance of the muon trigger chambers. In the electron analysis, the calorimeter transition region at $1.37<|\eta_e|<1.52$ is excluded.
The lower limit on the $\mt$ is imposed to further suppress background events that satisfy the lepton $\pt$ and $\mpt$ requirements.

In the muon channel, the background contribution from $\zmumu$ decays is suppressed by rejecting muons from opposite--charge pairs that have an invariant mass greater than 66~GeV.
These events are selected by requiring that one muon in the pair has $\pt>25\GeV$ and passes the quality requirements in Sect.~\ref{sec:mu-reco} and the other muon in the pair satisfies a lower $\pt$
threshold of $20\GeV$. 
In principle, this method allows for the possibility of accepting events with more than one $\Wboson$ boson. However, only one event in the data was found where two muons satisfy all signal selection requirements.  This selection vetoes 86\% of muons produced from $\Zboson$ bosons while retaining over 99\% of $\Wboson$ boson candidates. 
The 14\% of background muons that satisfy the selection criteria is attributable to instances where the second muon from the $\Zboson$ boson decay is produced outside the ID acceptance or has $\pt<20\gev$.

In the electron channel, the $\zee$ background contribution is suppressed by rejecting events with more than one electron satisfying the identification requirements from Sect.~\ref{sec:e-reco}. 
This selection retains over 99\% of signal events while rejecting 23\% of $\Zboson$ boson candidates. Events surviving the selection are attributable to instances where the second electron from the $\Zboson$ boson decay is either produced outside 
the ID acceptance~(26\%) or does not pass the relatively tight electron identification requirements~(74\%).

After applying all selection criteria, 3348 $\Wboson^{+}$ and 3185 $\Wboson^{-}$ candidates are detected in the muon channel. 
In the electron channel, 2893 $\Wboson^{+}$ and 2791 $\Wboson^{-}$ candidates are observed. 

\begin{figure}[!htbp]
\begin{center}
\resizebox{0.48\textwidth}{!}{
    \includegraphics[]{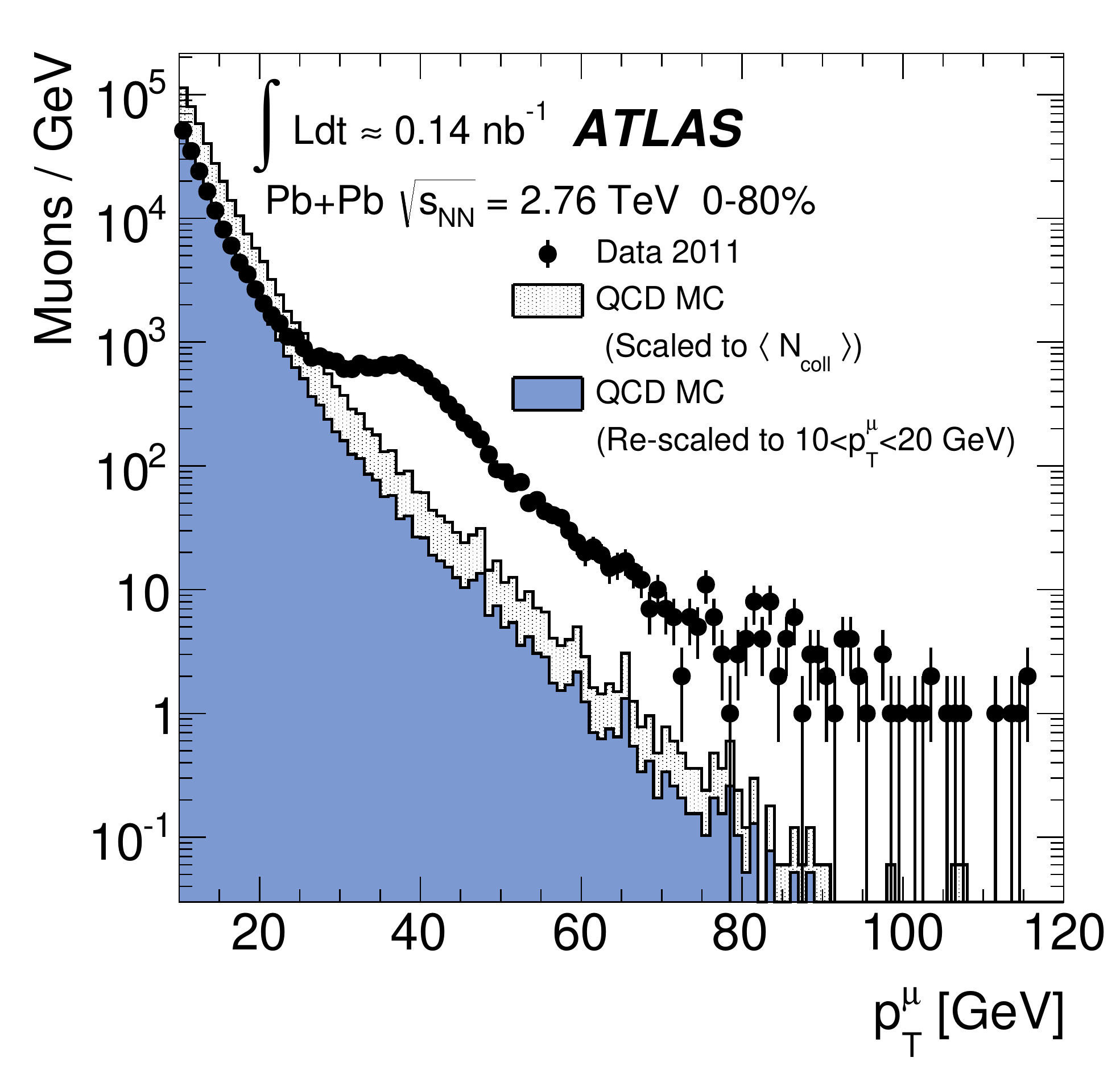}

}
\end{center}
\caption{ Muon transverse momentum distribution in the data~(points) before applying the signal selection requirements.  The $\pt$ distribution of QCD multi--jet processes from the MC simulation is also shown in the same figure. The shaded histogram is scaled to $\mNcoll$ and the solid histogram is rescaled to match the data in a control region $10<\pt^{\mu}<20 \GeV$. The background fraction from QCD multi--jet processes is determined from the number of muons in the MC surviving the final selection criteria.  
 }
\label{fig:qcd_mu}       
\end{figure}

\section{Background estimation}
\label{sec:bkg}

The main backgrounds to the $\wlnu$ channel arise from lepton production in electroweak processes and semileptonic heavy--flavour decays in multi--jet events. The former include $\wtaul$ events and $Z\rightarrow \ell^+\ell^-$ events, where one lepton from the $\Zboson$ boson is emitted outside the ID acceptance and produces spurious $\mpt$. Other sources of background that are considered include $\ztautau$ events, in which at least one tau decays into a muon or electron, and $\ttbar$ events, in which at least one top quark decays semileptonically into a muon or electron. These two background sources are negligible~($<$0.5\%) and are not taken into account in this analysis. 

\begin{figure*}[!hbtp]
\begin{center}
\resizebox{0.48\textwidth}{!}{
    \includegraphics[]{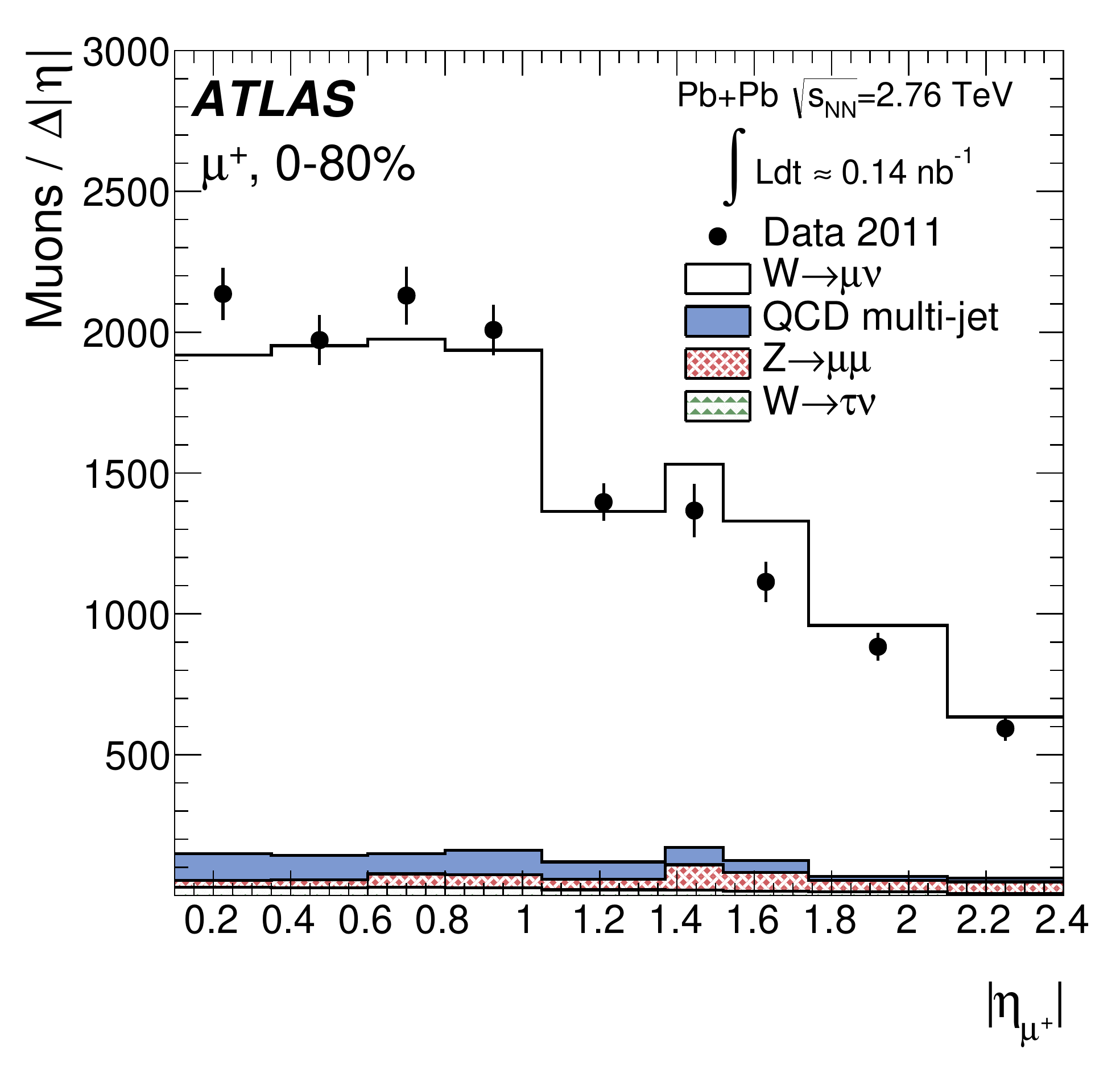}

}
\resizebox{0.48\textwidth}{!}{
    \includegraphics[]{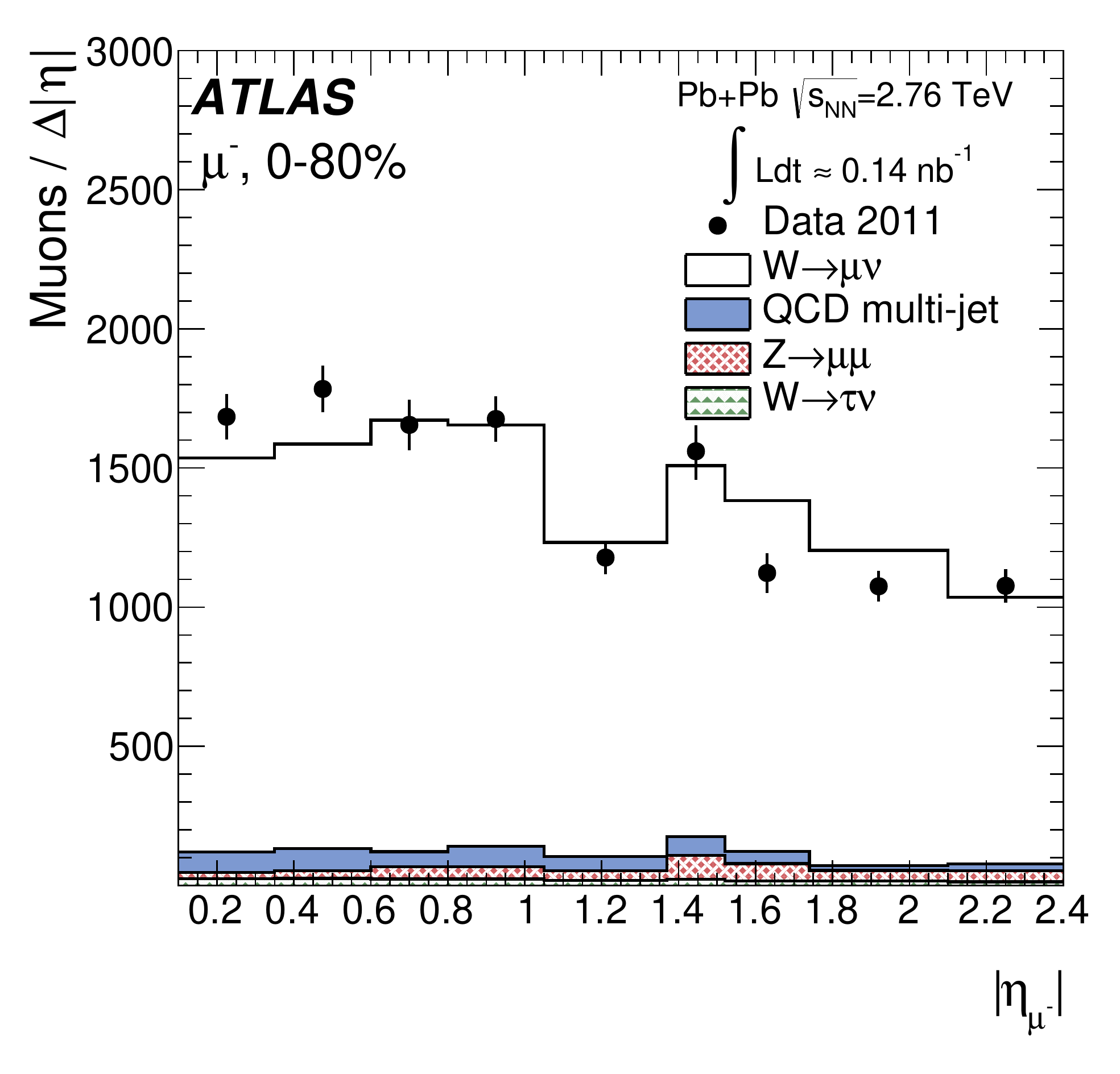}    
}

\resizebox{0.48\textwidth}{!}{
    \includegraphics[]{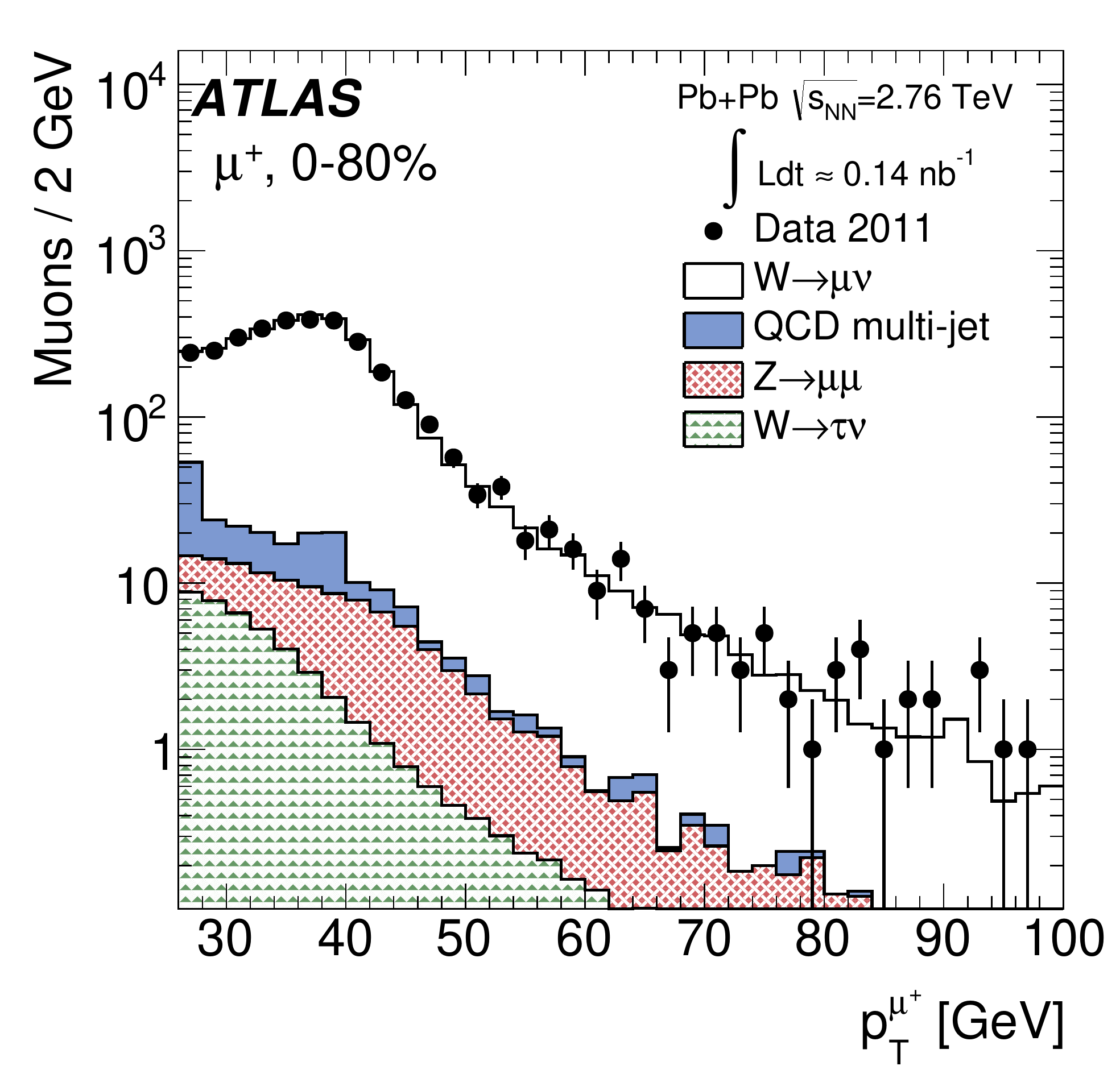}

}
\resizebox{0.48\textwidth}{!}{
    \includegraphics[]{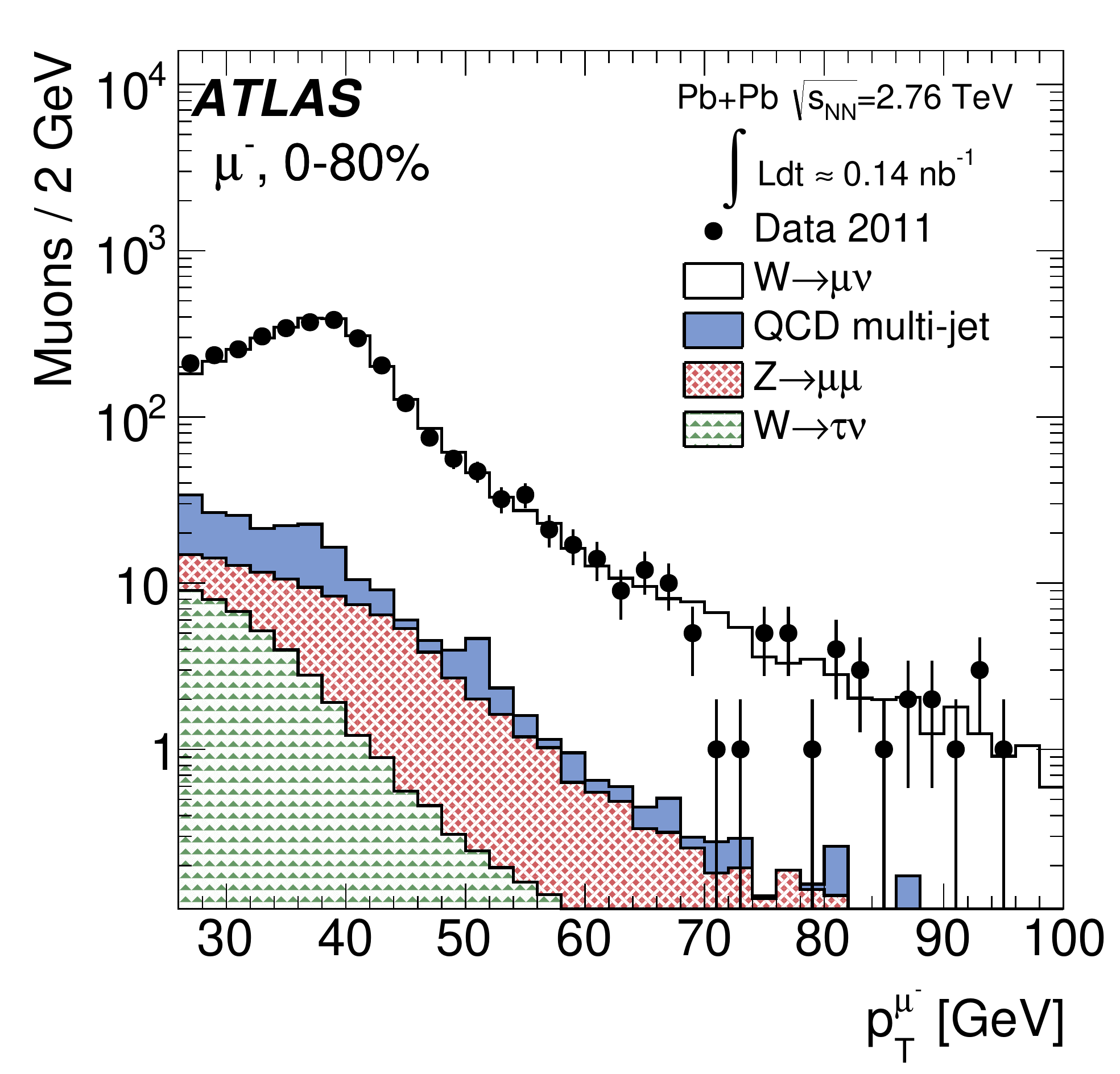}    
}

\end{center}
\caption{ Measured muon absolute pseudorapidity~(top) and transverse momentum~(bottom) distributions for $\wmunup$~(left) and $\wmunum$~(right) candidates after applying the complete set of selection requirements in the fiducial region, $\pt^\mu>25\gev, \mpt>25\gev, \mt>40\gev$ and $0.1<|\eta_\mu|<2.4$.
The contributions from electroweak and QCD multi--jet processes are normalised according to their expected number of events. The $\wmunu$ MC events are normalised to the number of background--subtracted events in the data. The background and signal predictions are added sequentially. }
\label{fig:1}       
\end{figure*}

\subsection{$\wmunu$ channel}

\begin{figure*}
\begin{center}
\resizebox{0.48\textwidth}{!}{
    \includegraphics[]{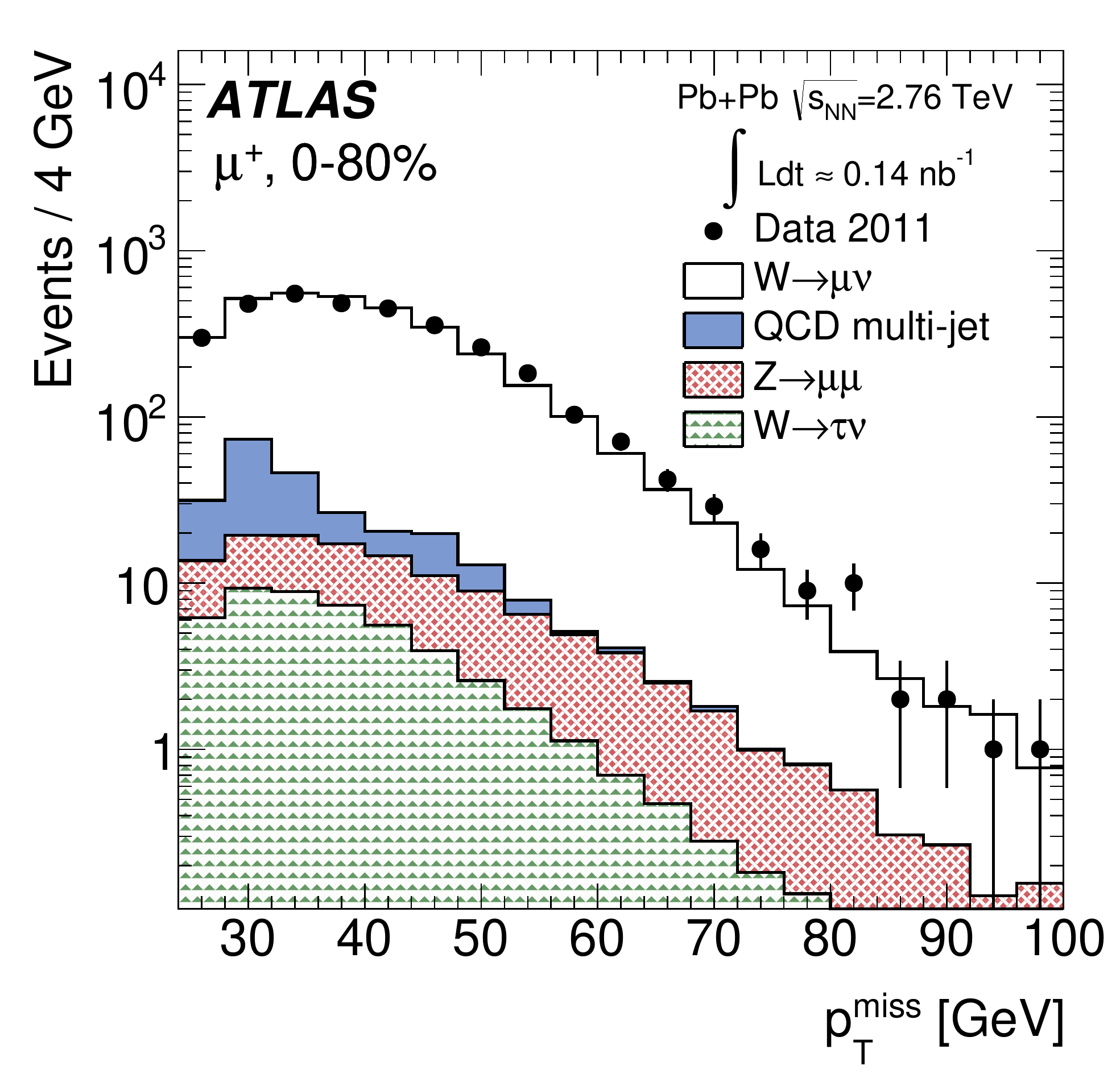}

}
\resizebox{0.48\textwidth}{!}{
    \includegraphics[]{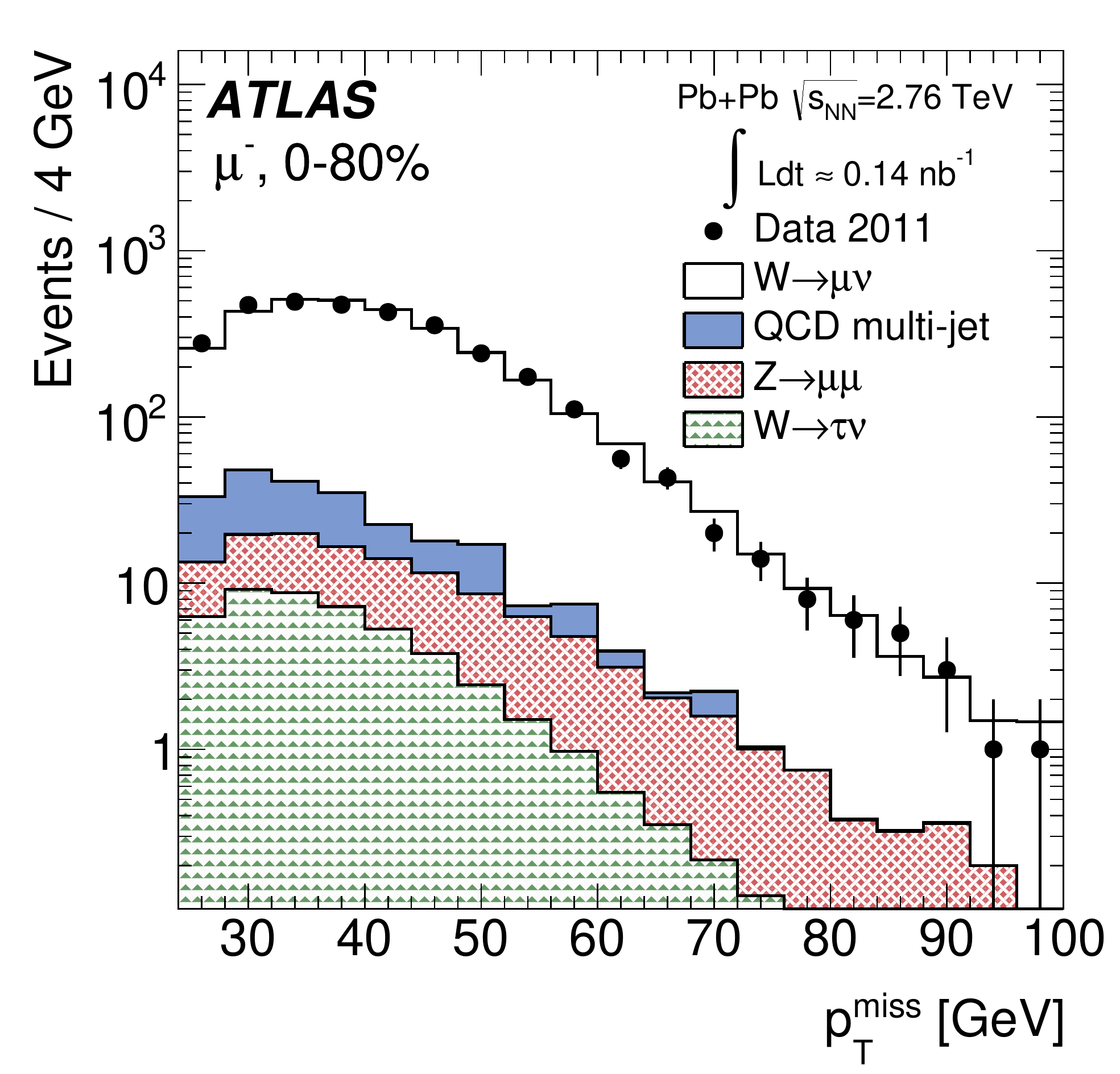}
}

\resizebox{0.48\textwidth}{!}{
    \includegraphics[]{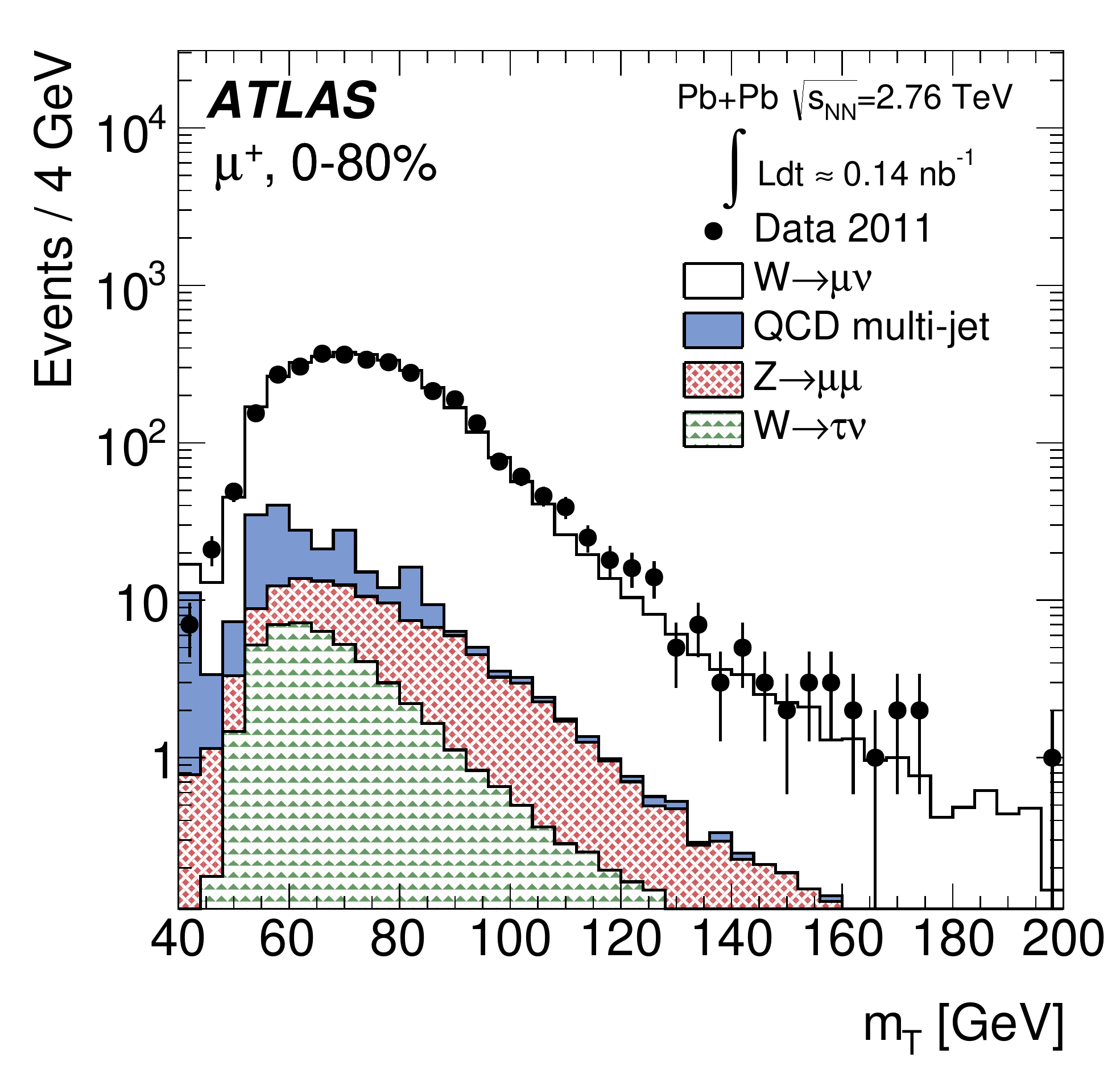}

}
\resizebox{0.48\textwidth}{!}{
    \includegraphics[]{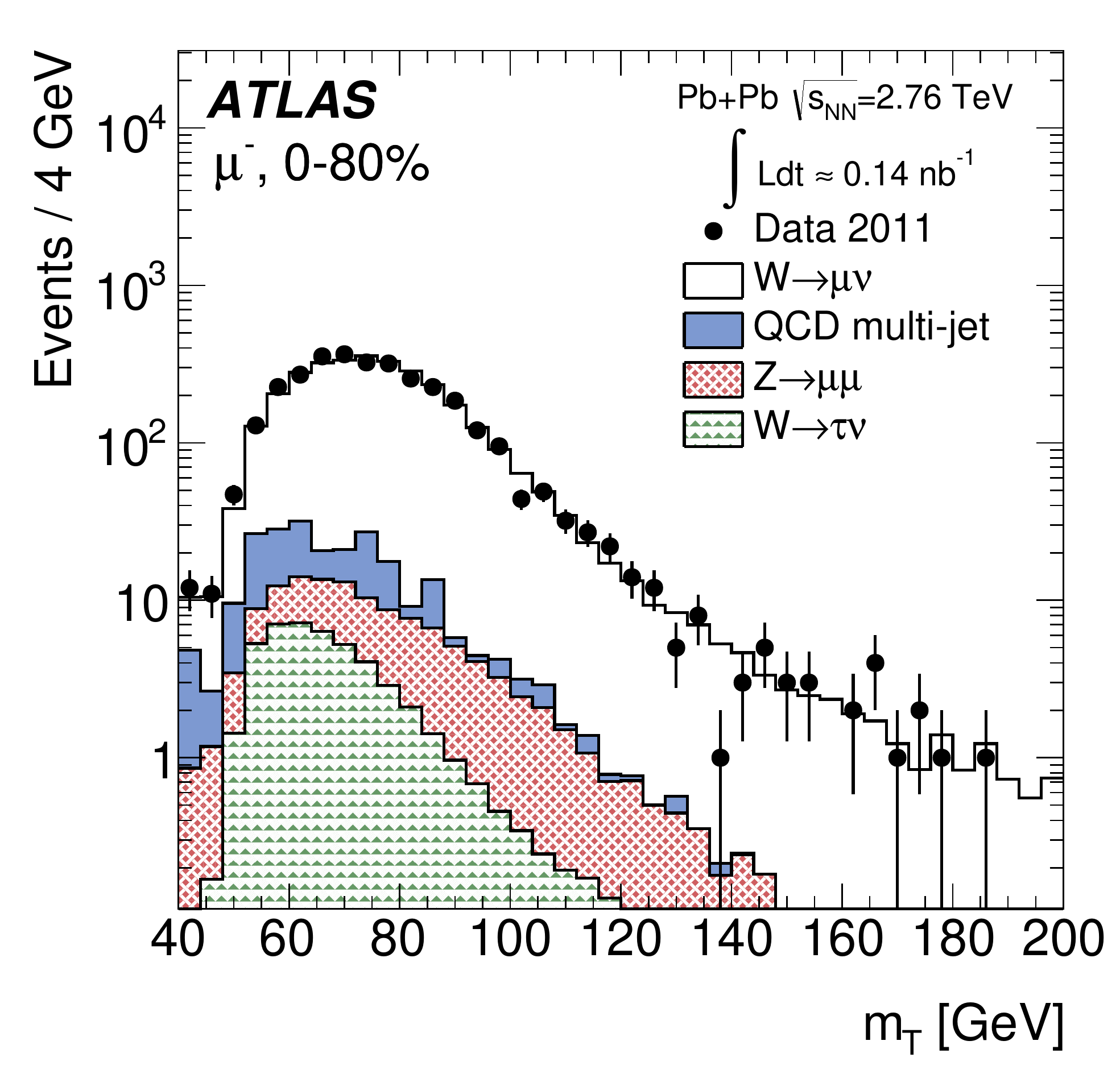}
}
\end{center}
\caption{ Measured missing transverse momentum~(top) and transverse mass~(bottom)  distributions for $\wmunup$~(left) and $\wmunum$~(right) candidates after applying the complete set of selection requirements in the fiducial region, $\pt^\mu>25\gev, \mpt>25\gev, \mt>40\gev$ and $0.1<|\eta_\mu|<2.4$. The contributions from electroweak and QCD multi--jet processes are  
normalised according to their expected number of events and added sequentially. The $\wmunu$ MC events are normalised to the number of background--subtracted events in the data. The background and signal predictions are added sequentially. }
\label{fig:3}       
\end{figure*}

\begin{figure}[!htbp]
\begin{center}

\resizebox{0.48\textwidth}{!}{
    \includegraphics[]{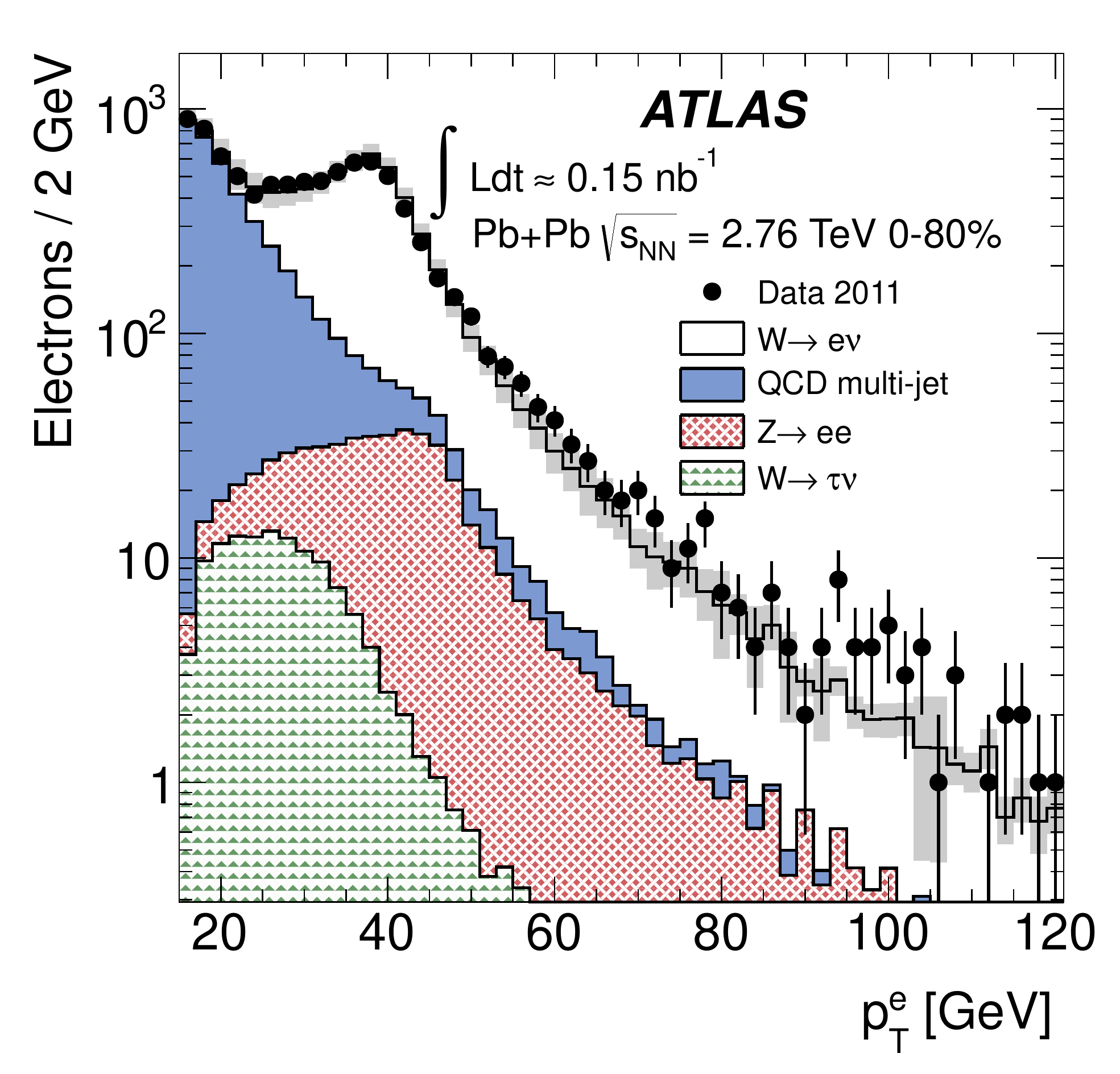}
}
\end{center}
\caption{ Electron transverse momentum distribution in the data~(points). The $\pt$ distribution of multi--jet events from a data control sample~(see text) and of simulated electroweak processes~($\wtaunu$ and $\zee$) are also shown. The total uncertainties from the fit are shown as solid grey bands.   
 }
\label{fig:qcd_el}      
\end{figure}

\begin{figure*}[!htbp]
\begin{center}
\resizebox{0.48\textwidth}{!}{
    \includegraphics[]{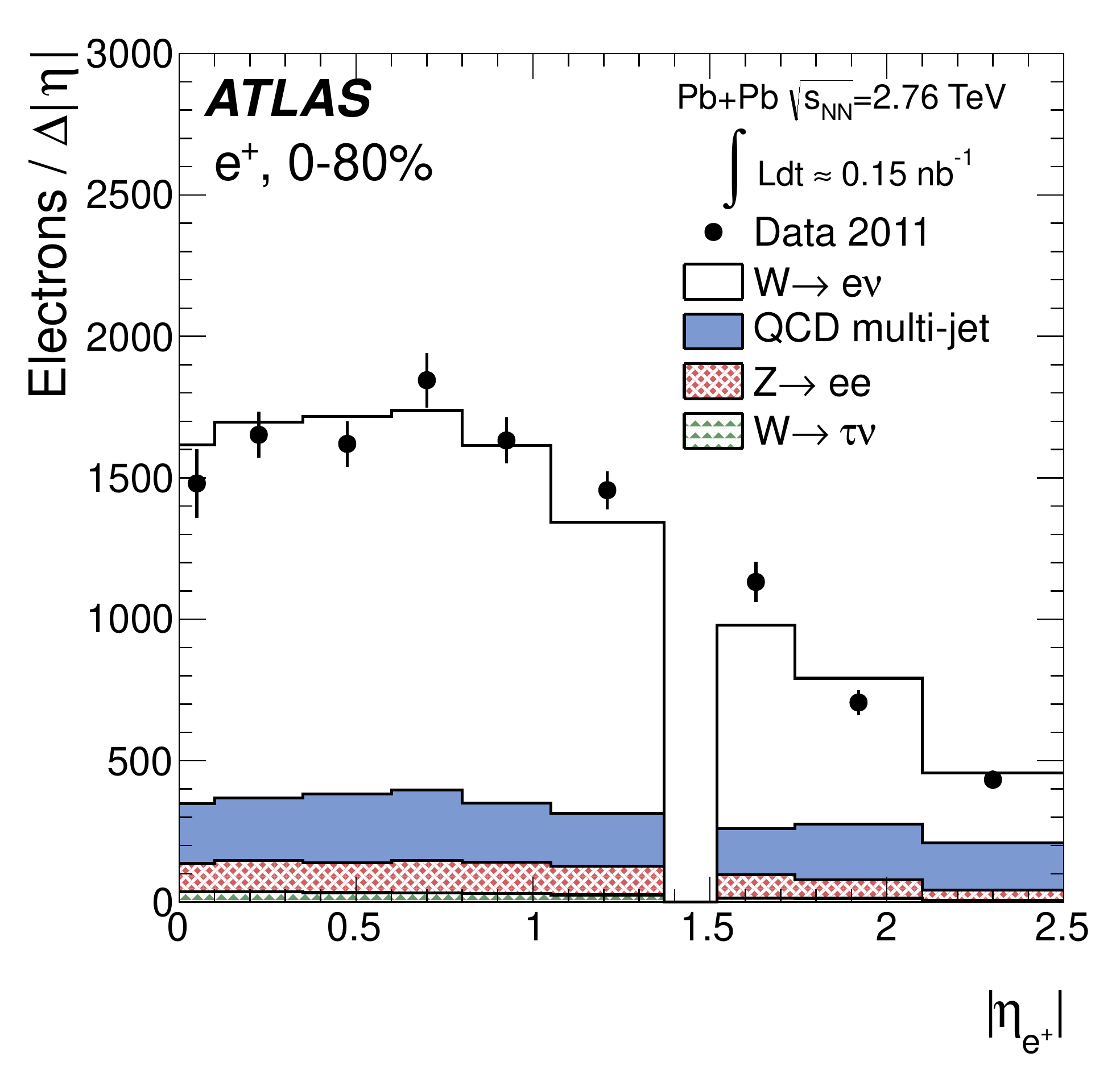}

}
\resizebox{0.48\textwidth}{!}{
    \includegraphics[]{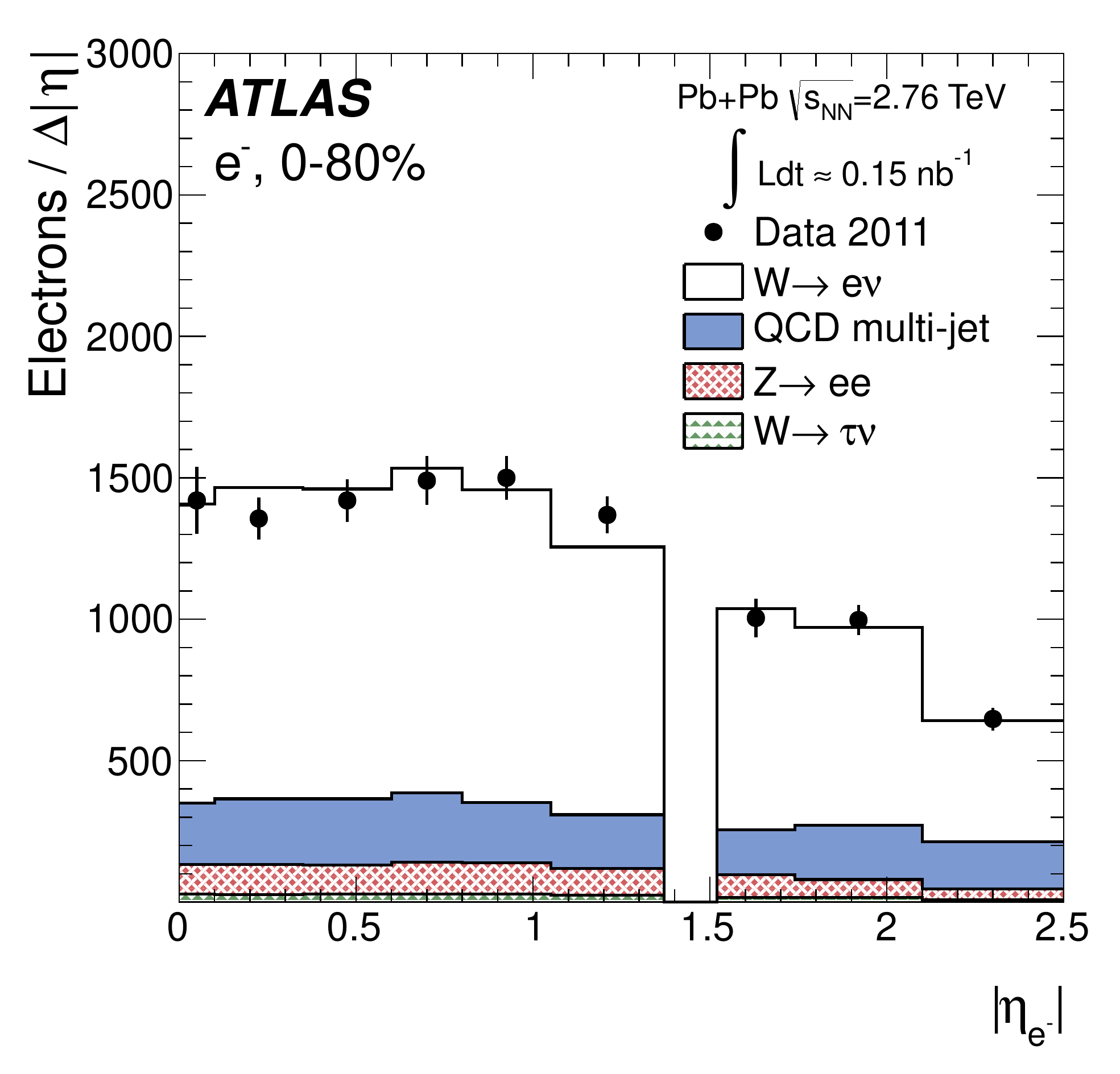}    
}
\resizebox{0.48\textwidth}{!}{
    \includegraphics[]{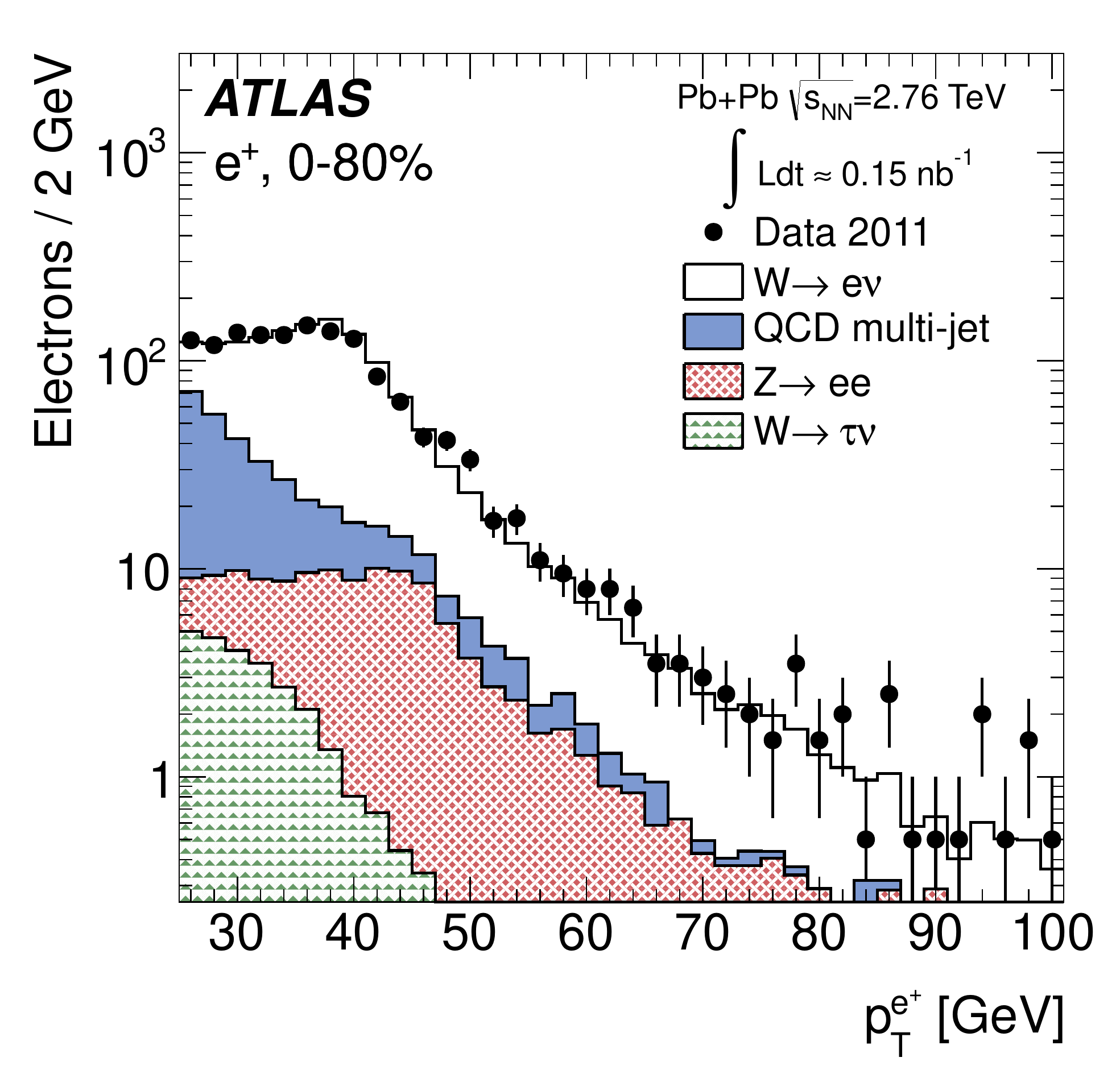}

}
\resizebox{0.48\textwidth}{!}{
    \includegraphics[]{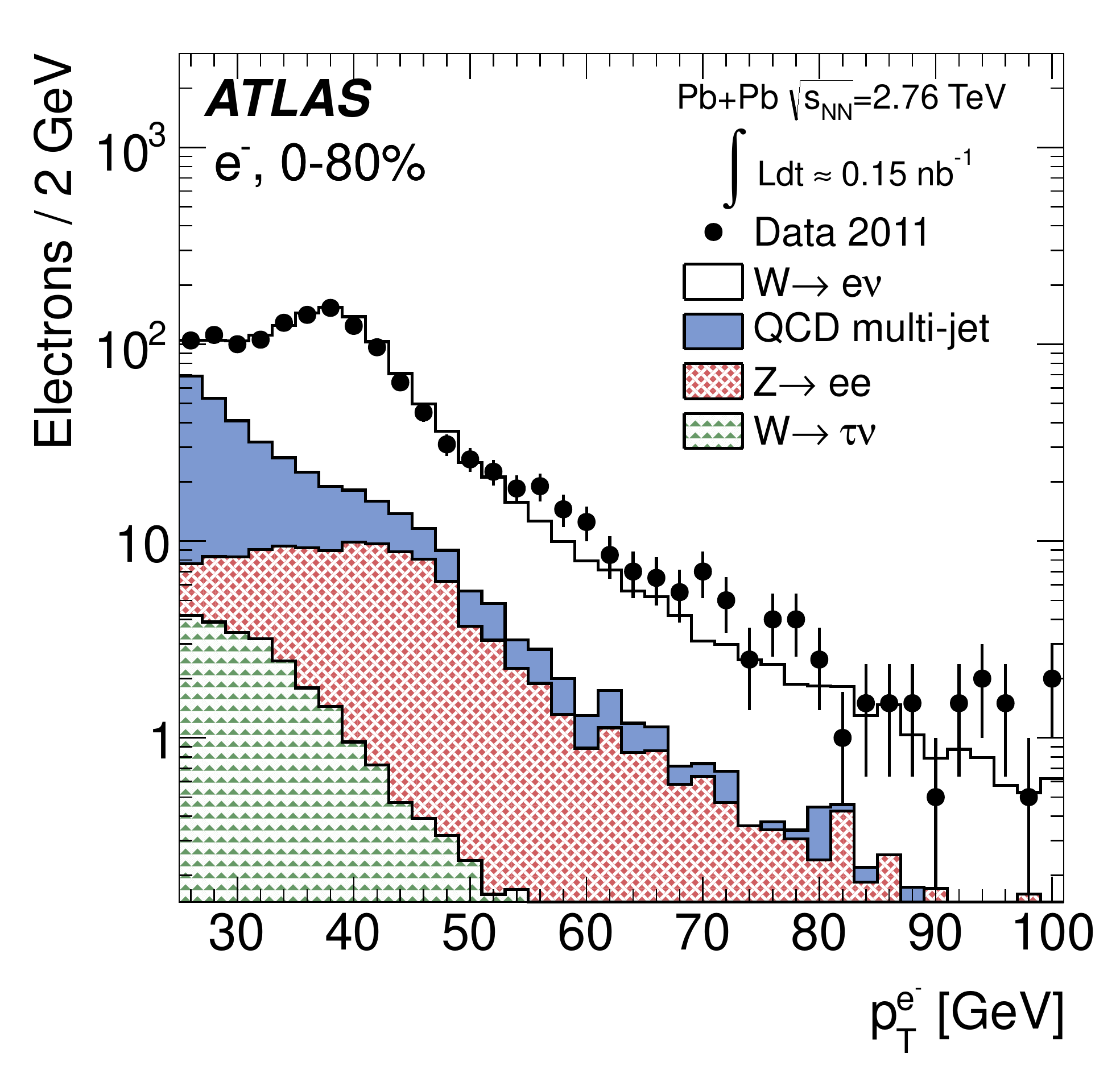}    
}
\end{center}
\caption{ Measured electron absolute pseudorapidity~(top) and transverse momentum~(bottom) distributions for $\wenup$~(left) and $\wenum$~(right) candidates after applying the complete set of selection requirements in the fiducial region, $\pt^e>25\gev, \mpt>25\gev, \mt>40\gev$ and $|\eta_e|<2.47$ excluding the transition region ~($1.37<|\eta_e|<1.52$). 
The contributions from electroweak and QCD multi--jet processes are normalised according to their expected number of events. The $\wenu$ MC events are normalised to the number of background--subtracted events in the data. The background and signal predictions are added sequentially.}
\label{fig:e-eta}       
\end{figure*}

\begin{figure*}[!hbtp]
\begin{center}
\resizebox{0.48\textwidth}{!}{
    \includegraphics[]{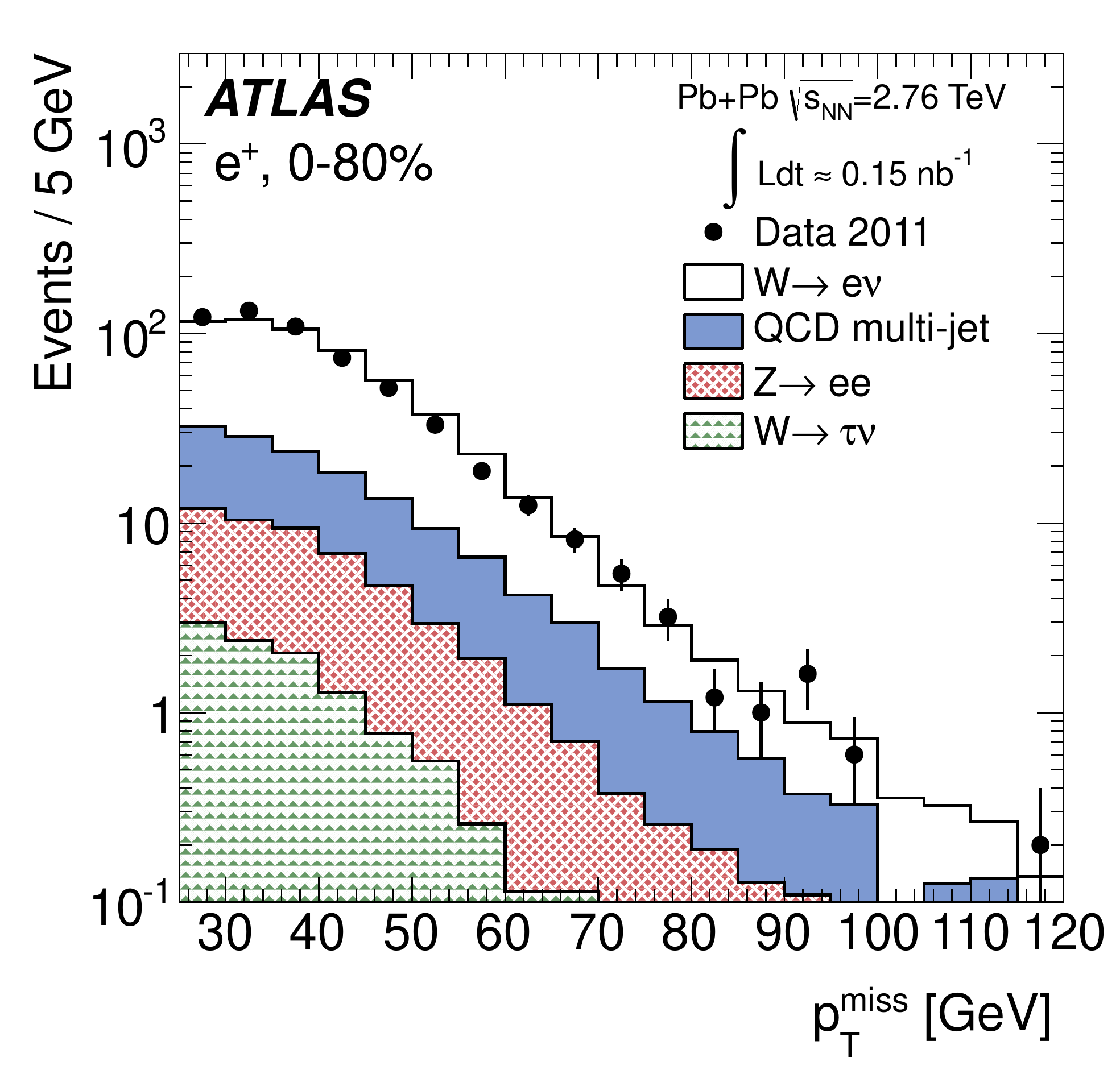}

}
\resizebox{0.48\textwidth}{!}{
    \includegraphics[]{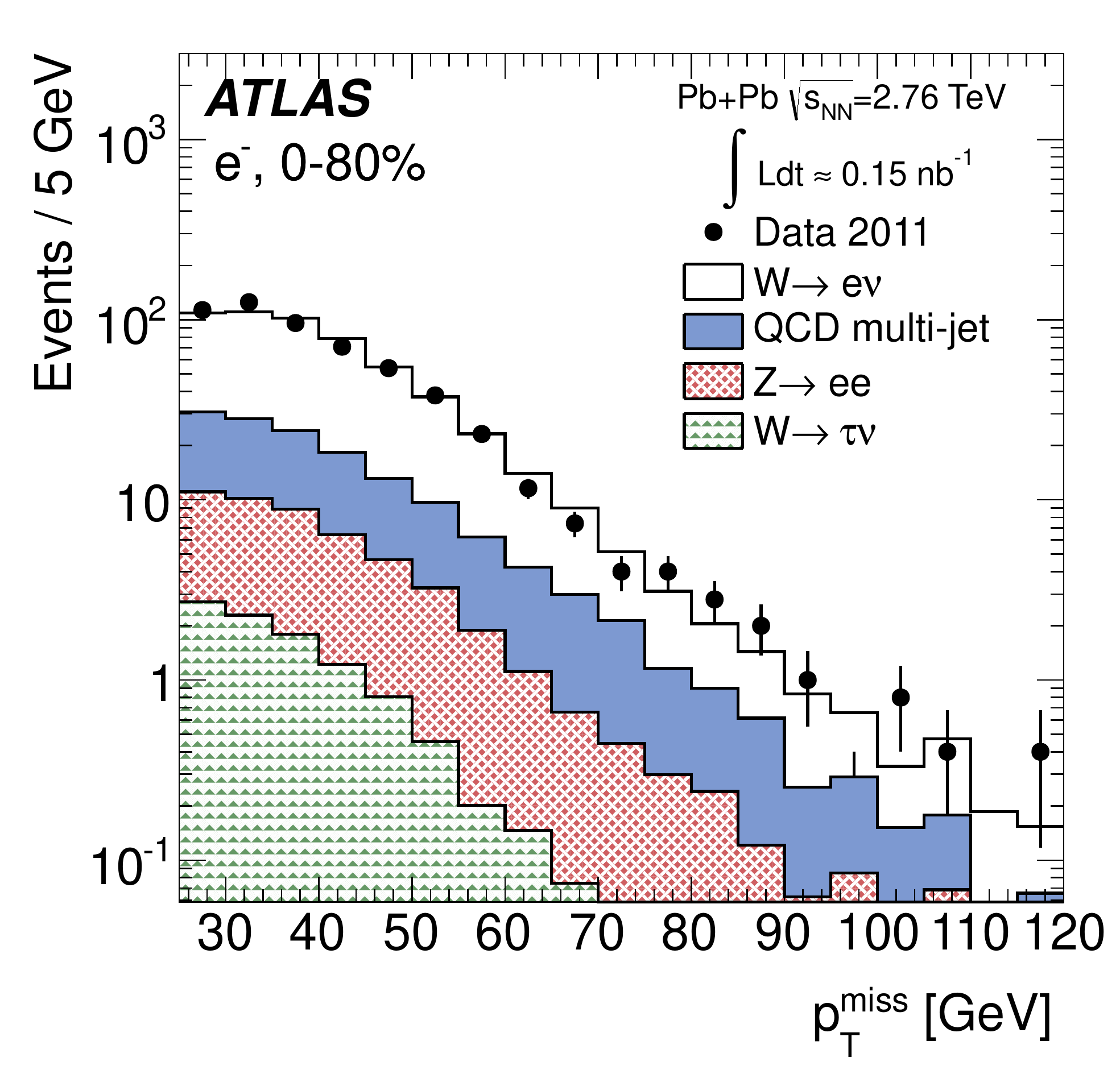}
}
\resizebox{0.48\textwidth}{!}{
    \includegraphics[]{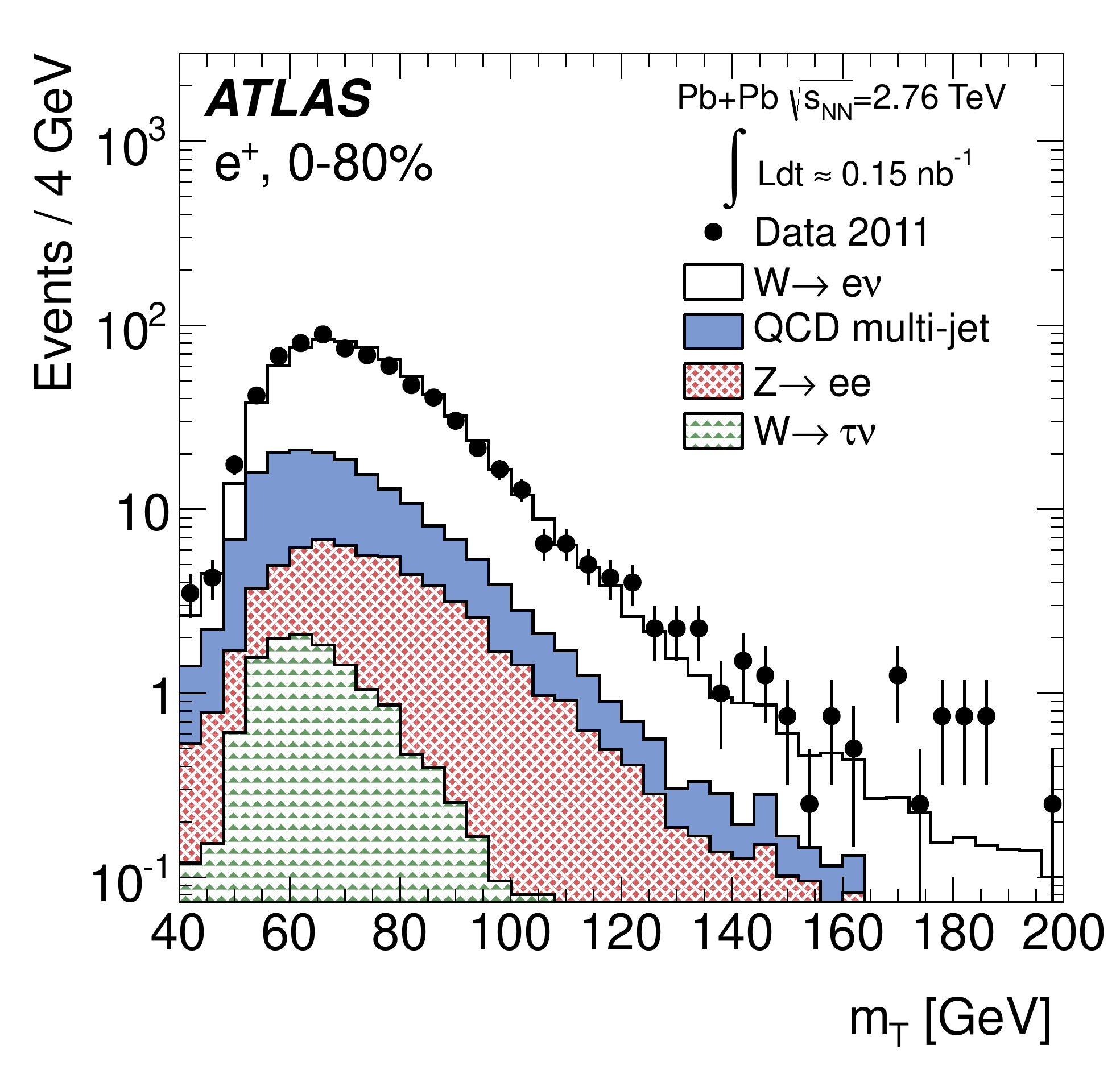}

}
\resizebox{0.48\textwidth}{!}{
    \includegraphics[]{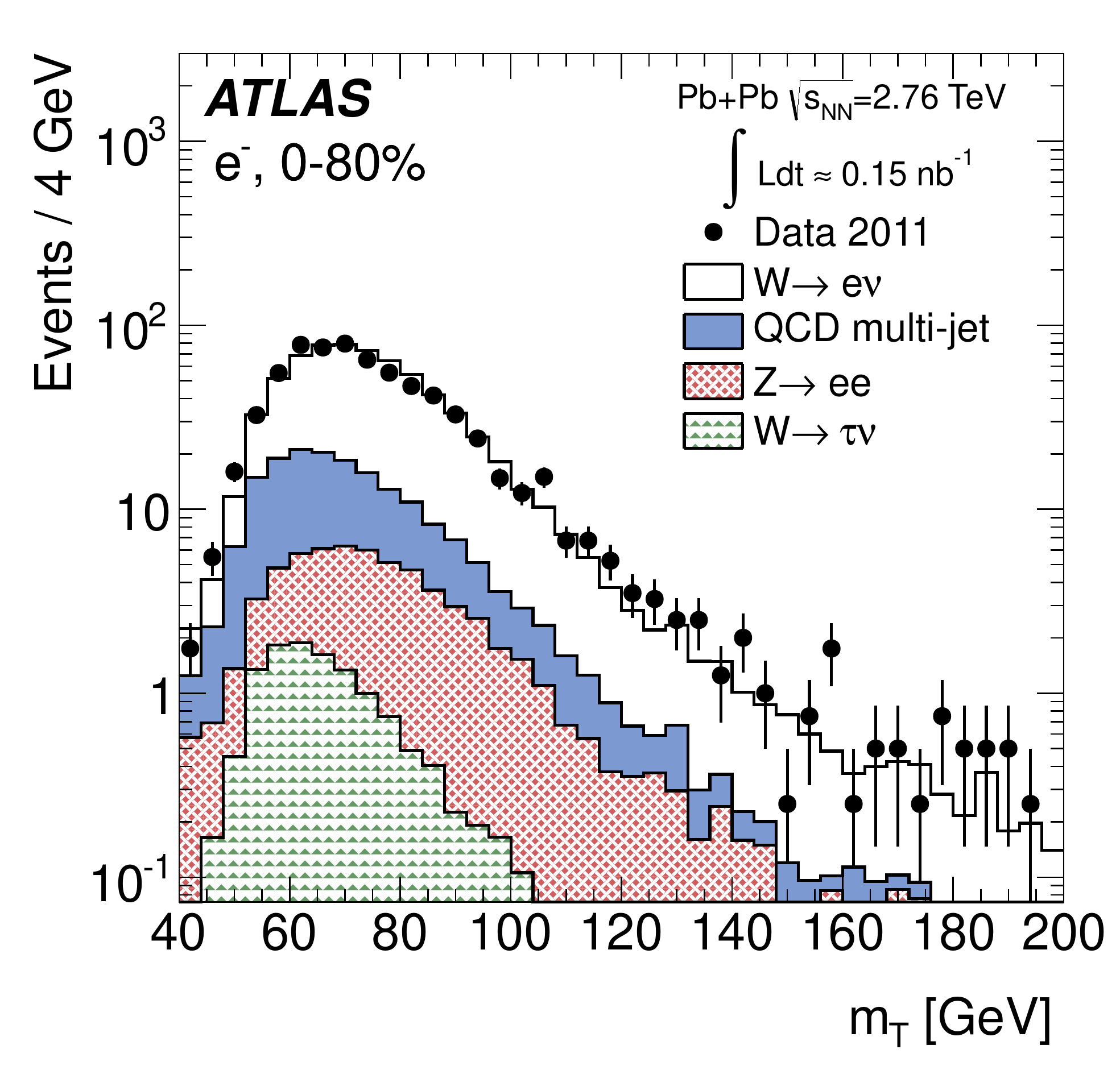}
}
\end{center}
\caption{ Measured missing transverse momentum~(top) and transverse mass~(bottom) distributions for $\wenup$~(left) and $\wenum$~(right) candidates after applying the complete set of selection requirements in the fiducial region, $\pt^e>25\gev, \mpt>25\gev, \mt>40\gev$ and $|\eta_e|<2.47$ excluding the transition region ~($1.37<|\eta_e|<1.52$). The contributions from electroweak and QCD multi--jet processes are normalised according to their expected number of events. The $\wenu$ MC events are normalised to the number of background--subtracted events in the data. The background and signal predictions are added sequentially. }
\label{fig:e-met}       
\end{figure*}

In the muon channel, the total number of background events from QCD multi--jet processes is estimated using a partially data--driven method.
The dijet muon yields per $\PbPb$ event in the MC simulation are normalised to the $pp$ cross--section and scaled by the number of binary collisions and $\PbPb$ events in the data.
The resulting distribution is represented by the shaded histogram in Fig.~\ref{fig:qcd_mu}. To take into account jet energy--loss in the medium, the MC distribution is rescaled to the data in a control region dominated by QCD multi--jet events in the range $10<\pt^{\mu}<20\GeV$~(solid histogram). 
This scale factor is on average $0.4$ over all $|\eta_{\mu}|$ intervals and centrality classes. 
As a cross--check, the shape of the rescaled QCD multi--jet background distribution 
was compared to that of a control sample consisting of anti--isolated muons from the data. They are found to agree well, 
confirming that the distributions in Fig.~\ref{fig:qcd_mu} are an accurate representation of the multi--jet background in the data.
The number of expected QCD multi--jet events is determined by extrapolating the rescaled MC distribution from the control region to the signal $\pt^{\mu}$ region above $25\GeV$.  
The fraction of background events in the data is then calculated from the 
ratio of the number of QCD multi--jet events surviving final selection in the MC and the number of $\Wboson$ candidates in the data. 
This is performed as function of $\eta_{\mu}$ and centrality.
The background fraction is also determined separately for $\mu^{+}$ and $\mu^{-}$, and no charge dependence is observed. 
The multi--jet background fraction is estimated to be on average 3.7\% of the total number of $\Wboson^{\pm}$ boson candidates, varying from 2.0\% to 5.4\% as a function of $\eta_{\mu}$ and centrality.

The estimated number of background events from electroweak processes is determined separately for the $\zmumu$ and $\wtaunu$ channels.
The background from $\zmumu$ events is determined in each $\eta_{\mu}$ interval from MC simulation and 
scaled to reproduce the actual number of $\zmumu$ events observed in the data~\cite{Aad:2012ew} in each centrality class.
This contribution is on average 2.4\% relative to the total number of $\Wboson$ boson candidates and ranges from 1.0\% at central $|\eta_{\mu}|$ to 3.2\% in the forward region. 
Background events originating from $\wtaumu$ decays are estimated by calculating the ratio of the number of $\wtaumu$ and $\wmunu$ events that satisfy the analysis selection in the 
simulation. This fraction is on average 1.5\% in each $|\eta_{\mu}|$ interval and centrality class and is applied to the number of observed signal candidates. 
Variations between bins are at the level of 1.3--1.8\%.   
The expected background from all sources in the $\wmunu$ channel amounts to 7.6\% of the total number of $\Wboson$ boson candidates. 

Figure~\ref{fig:1} shows the $|\eta_{\mu}|$ and $\pt^{\mu}$ distributions for positively and negatively charged
muons after final event selection.
Figure~\ref{fig:3} presents the event $\mpt$ and $\mt$ distributions.
In each figure, the data are compared to signal and background distributions from MC simulation in the same phase space.
The background distributions are normalised to the expected number of events, whereas the 
signal MC distribution is normalised to the number of background--subtracted events in the data. The background and signal 
predictions in Figs.~\ref{fig:1} and~\ref{fig:3} are added sequentially, beginning with the contribution from $\wtaunu$.  

\subsection{$\wenu$ channel}

A partially data--driven method is used to estimate the QCD multi--jet background observed in $\wenu$ candidate events. 
This method involves using a control sample from the data to construct a QCD background template and simulated $\wenu$ events to construct a signal template. The control sample is selected by employing looser electron identification criteria based solely on shower shape information and inverting the isolation requirement. In addition, if the event contains a jet reconstructed at EM scale with $\et>25\gev$, the difference between the azimuthal angle of the jet and $\mpt$ is required to be greater than $\pi/2$. This condition suppresses events with spurious $\mpt$ originating from miscalibration of a jet~\cite{Aad:2011mk}. The nominal $\mpt$ and $\mt$ criteria are also applied to the control sample. 
The background and signal templates are fit to the data as a function of $\pt^e$ in the signal region after electroweak background subtraction. 
A result of the fit is shown in Fig.~\ref{fig:qcd_el}. 
The fit result slightly underestimates the data at $\pt^e\simeq60$~\GeV, but this difference is within the total uncertainty of the fit. 
A significant contribution to this uncertainty comes from the limited number of events available for determining the QCD multi--jet background.
The fitting is performed in all centrality bins and results in a total background estimation of 16.7\% of $\wenu$ candidate events in the 0--80\% centrality class. As in the muon channel, this background fraction is charge--independent. 

The background from electroweak processes with electrons in the final state is estimated from the MC samples listed in Table~\ref{tab:mcsamples}. The nominal selection criteria of this analysis are imposed on each MC sample.  The absolute normalisation is derived from the $\Wboson$ and $\Zboson$ \powheg\ cross--sections in \pp\ collisions. These cross--sections are scaled by $\mNcoll$ in each centrality bin and normalised to the integrated luminosity of the \PbPb\ data sample.  This method gives a valid estimate of the electroweak background in this analysis since ATLAS has recently demonstrated that the $\zee$ yields in \PbPb\ collisions at \energy\ are consistent with the \pp\ expectation scaled by $\taa$ to within 3\%~\cite{Aad:2012ew}. The $\zee$ background is the dominant electroweak background in this analysis and amounts to 6.5\% of the total $\wenu$ candidate events. 
The background from $\wtaunu$ contributes an additional 2.5\%. Electrons from $\ztautau$ and $t\bar{t}$ are found to be $<$0.3\% and $<$0.1\%, respectively. As with the muon channel, the latter two background sources are considered negligible.

Figure~\ref{fig:e-eta} shows the $|\eta_e|$ and $\pt^e$ distributions for positively and negatively charged
electrons after final event selection.
Figure~\ref{fig:e-met} presents the event $\mpt$ and $\mt$ distributions.
In each figure, the data are compared to signal and background distributions from MC simulation in the same phase space.
The background distributions are normalised to the expected number of events, whereas the 
signal MC distribution is normalised to the number of background--subtracted events in the data. The background and signal 
predictions in Figs.~\ref{fig:e-eta} and~\ref{fig:e-met} are added sequentially, beginning with the contribution from $\wtaunu$. 

\section{Yield correction procedure}
\label{sec:correction}

In order to correct the data for losses attributable to the trigger, reconstruction, and selection efficiencies, a correction factor is applied to the measured yields after background subtraction. 
This correction factor $C_{\Wboson^{\pm}}$ is defined by the following ratio: 

\begin{equation}
C_{W^{\pm}} = \frac{N_{W}^{\mathrm{rec}}}{N_{W}^{\mathrm{gen,fid}}},
\label{eqnCw}
\end{equation}

\noindent where $N_{W}^{\mathrm{rec}}$ represents the number of $\wlnu$ events reconstructed in the fiducial region and satisfying final selection criteria, and $N_{W}^{\mathrm{gen,fid}}$ signifies the number of $\wlnu$ events in the same phase space at the generator--level.
This is calculated separately for each charge, $|\eta_{\ell}|$ interval, and centrality class.
The denominator in Eq.~(\ref{eqnCw}) is evaluated directly from the boson decay i.e. Born level; this way of constructing the correction factor accounts for effects due to migration and QED radiation in the final state.
Corrections for reconstruction and selection are derived solely from the signal MC simulation, whereas the trigger efficiencies are obtained from the data in each $|\eta_{\ell}|$ interval and centrality class. 

In both the muon and electron channels, the $C_{W^{\pm}}$ significantly depends on the event centrality and $|\eta_\ell|$. 
In the muon channel, the integrated $C_{W^{\pm}}$ is $(67.4\pm0.2)\%$, ranging from 32\% in the most central events in the highest $|\eta_{\mu}|$ region to 85\% in the most peripheral events at mid-pseudorapidity. 
In the electron channel, the integrated $C_{W^{\pm}}$ is $(39.2\pm0.3)\%$, ranging from 34\% in the most central events to 51\% in the most peripheral centrality class.
The large variations in the $C_{W^{\pm}}$ are attributable to two main factors: areas of the detector with limited coverage and the centrality dependence of the isolation efficiency and $\mpt$ resolution. 

The differential $\Wboson$ boson production yields in the fiducial region are computed as:

\begin{equation}
N_{\Wboson^{\pm}}(|\eta_\ell|,\mathrm{centrality}) = \frac{N_{\Wboson^\pm}^{\mathrm{obs}}-N_{}^{\mathrm{bkg}}}{C_{\Wboson^{\pm}}},
\label{eqCS}
\end{equation}

\noindent where \emph{$N^{\mathrm{obs}}_{\Wboson^\pm}$} signifies the number of candidate events observed in the data and \emph{$N^{\mathrm{bkg}}_{}$} the number of background events in a given $|\eta_{\ell}|$ and centrality class. 

The combination of the results from each channel are reported both as an integrated result in each centrality class and as a differential measurement as a function of $|\eta_\ell|$.
The integrated result requires the extrapolation of each measurement to the full pseudorapidity region,~$|\eta_\ell|<2.5$ -- this includes the excluded regions discussed above. Correction factors for this 
extrapolation are derived from the signal MC simulation and increase the integrated yield for muons by 7.5\% and electrons by 6.6\%. In the differential measurement as a function of $|\eta_\ell|$,
the extrapolation is performed only in the most forward bin up to $|\eta_\ell|=2.5$. The correction increases the number of signal candidates in this bin by 28\% in the muon channel and 7\% in the electron channel.

\section{Systematic uncertainties}
\label{sec:syst}
The systematic uncertainties are studied separately for each charge, $|\eta_\ell|$, and centrality class.
The magnitude by which each uncertainty is correlated from bin--to--bin is determined from 
the change in the corrected yields as a function of $|\eta_\ell|$ and centrality after applying a systematic variation.
The sources of uncertainty considered fully correlated between bins are as follows: the $\mpt$ resolution, electroweak and QCD multi--jet background estimations, lepton isolation efficiencies, 
lepton and track reconstruction efficiencies, lepton energy/momentum scales and resolutions, extrapolation corrections and $\mNcoll$.
The dominant systematic uncertainty in both channels 
originates from the missing transverse momentum resolution.
In the asymmetry and charge ratio measurements, 
uncertainties correlated between charges largely cancel.  
This correlation is determined for each source of systematic uncertainty from the variation in the charge ratio measurements with respect to the nominal values. 
\subsection{Muon channel}

The resolution on the $\mpt$ (described in Sect.~\ref{sec:event-sel}) worsens 
with an increasing soft particle contribution to the vector sum of Eq.~(\ref{eqnMPT}). This in turn depends on the 
lower track $\pt$ threshold. 
The variation in the resolution with lower track $\pt$ threshold is attributable  
to sources of spurious $\mpt$ -- e.g. undetected tracks, limited detector coverage, inactive material, finite detector resolution. These sources become amplified when 
a larger number of tracks are considered in the vector sum.
A larger $\sigma_{\mathrm{miss}}$ in the $\mpt$ distribution implies a larger uncertainty of the true neutrino $\pt$. 
However, setting a lower track $\pt$ threshold too high can also introduce sources of fake $\mpt$ by vetoing tracks required to 
balance the transverse energy of the event. Therefore, to optimise the $\mpt$ calculation, several lower track $\pt$ thresholds were studied in MB events and 
3~\GeV\ is considered optimal. 
To quantify the uncertainty on the optimisation, the $\pt$ threshold of the tracks used in Eq.~(\ref{eqnMPT}) is varied in both data and MC simulation by $\pm 1\GeV$ relative to the nominal track $\pt$ threshold. 
All background sources, correction factors, and signal yields are recalculated during this procedure, resulting in an estimated uncertainty in the signal yield of 2.0--4.0\%.

The uncertainty in the QCD multi--jet background estimation arises primarily from the extrapolation procedure. There are two contributing factors: 
how well the MC simulation represents the shape of the QCD multi--jet muon $\pt$ distribution -- particularly in the high--$\pt$ region -- and to what degree this distribution is altered by jet energy--loss in the medium. 
Both contributions may be accounted for by scaling the muon $\pt$ distribution from simulated QCD multi--jet events by a $\pt$--dependent nuclear modification factor. The scale factors are calculated according to the procedure from Ref.~\cite{CMS:2012aa} and are defined as the ratio of the inclusive charged hadron yield per binary collision in a heavy--ion event
and the charged hadron yield in a \pp\ collision. This is performed for each centrality class.
Since there is little difference between the nuclear modification factor between heavy--flavour muons and inclusive charged hadrons~\cite{CMS:2012aa,Abelev:2012qh}, this 
scaling procedure is a valid estimation of the extrapolation uncertainty. 
Applying this factor to each muon $\pt$ bin results in a maximum uncertainty in the QCD multi--jet background of 50\% and variations in the final signal yields from 0.4\% to 2.0\%.

The electroweak background uncertainty is estimated separately for $\zmumu$ and $\wtaunu$.
The uncertainty in the $\Zboson$ boson background estimation is determined by scaling the number of $\Zboson$ events in each $\eta_{\mu}$ interval to the number of events estimated from the MC simulation rather than those observed in the data in each centrality class.
The variation in the number of $\wmunu$ events in each $|\eta_\mu|$ or centrality class with respect to the nominal yields is $<0.1\%$.
The systematic error in the $\tau$ background estimation is evaluated by 
assuming that the muon selection efficiencies for the $\mpt$ and $\mt$ requirements in the $\wtaumu$ sample are 
identical to those in the $\wmunu$ sample for muons with $\pt^{\mu}>25\GeV$.
Estimating the $\tau$ background with these efficiencies from the $\wmunu$ sample results in a variation in the signal yields no larger than $0.1\%$ of the nominal number of signal events in the data. 
Other sources of background from $\ztautau$ and $\ttbar$ events are also included as a systematic uncertainty and result in a signal variation of less than $0.2\%$.

A systematic uncertainty attributable to the modelling accuracy of the isolation in the MC simulation is assessed by varying the $\Delta R$ and $\sum\pt^{\mathrm{ID}}$ requirements in both data and simulation.
This uncertainty is estimated by re--evaluating the yields either with a larger $\Delta R$ or a larger $\sum\pt^{\mathrm{ID}}$.
The $\Delta R$ around the muon momentum direction is increased from 0.2 to 0.3, and the requirement on the $\sum\pt^{\mathrm{ID}}$ is increased from 10\% to 20\% of the muon $\pt$. 
This results in a yield variation of 1--2\% in each centrality, $|\eta_{\mu}|$, or charge class. 

Systematic uncertainties related to the $C_{\Wboson^\pm}$ correction originate from uncertainties in the muon $\pt$ resolution, reconstruction efficiency, and trigger efficiency. These uncertainties were previously 
evaluated for the 2011 heavy--ion data--taking period in Ref.~\cite{Aad:2012ew}. A short summary of the methodology used in estimating these uncertainties and their respective contributions to the $\Wboson$ analysis is provided below. 
An uncertainty in the muon $\pt$ resolution due to differences in the detector performance in simulation relative to actual data--taking conditions 
is estimated by additionally smearing the $\pt$ of muons in the MC simulation in the range allowed by the systematic uncertainties in 
Ref.~\cite{ATLAS-CONF-2013-088}. 
The correction factors are then re--evaluated, 
and the yield variation is used as the systematic uncertainty. 
The relative uncertainty from this procedure results in a variation of less than 1.0\% in the number of signal events in each $\eta_\mu$, centrality, and charge class. 
Uncertainties in the muon reconstruction efficiency are also estimated from $\zmumu$ events.
To estimate this uncertainty, $\zmumu$ MC events are re--weighted such that the ratio of the number of muon pairs reconstructed using both 
the ID and MS components and muon pairs reconstructed using only the MS component -- with no restriction on the ID component -- agree in data and the MC simulation. 
The reconstruction efficiencies in the MC simulation are then recalculated and result 
in an additional 1.0\% uncertainty in the number of $\wmunu$ events.
Uncertainties in the muon trigger efficiency are determined from differences in the efficiencies 
calculated using single muons from MB events and a tag--and--probe method applied to a $\zmumu$ sample.
This results 
in yield variations of 0.4\%.  

Scaling uncertainties in $\mNcoll$ are also applied when reporting the yields per binary collision. These were shown in Table~\ref{tab:centrality} and 
arise from possible contamination due to photonuclear events and diffractive processes. The procedure for calculating these 
uncertainties is described in detail in Ref.~\cite{Miller:2007ri}. This uncertainty is largest in the most peripheral events and amounts to 9.4\%.
Integrated over all events the $\mNcoll$ uncertainty is around 8.5\%.

The extrapolation of the yields over $|\eta_\mu|<2.5$ also introduces a source of systematic uncertainty. This uncertainty is mainly attributable to the PDF uncertainty, 
which has been studied extensively in \pp\ collisions at the LHC by ATLAS~\cite{Aad:2011dm} using the same PDF set that this analysis uses to correct the data. 
The uncertainties are derived from differences in the correction factor using various PDF sets, differences due to the parton-shower modelling, and the PDF error eigenvectors. These individual contributions are added in quadrature and result in uncertainties at the 0.2\% level. An uncertainty of 0.3\% is associated with the differential production measurement in the highest $|\eta_\mu|$ bin.

Table~\ref{tab:muon-syst} presents a summary of the 
maximum values for all systematic uncertainties included in the muon channel. 
Systematic uncertainties correlated between different centrality or $|\eta_{\mu}|$ intervals are 3--5\%.   
The bin--uncorrelated systematic uncertainties, which are comprised of statistical uncertainties from the background estimation, trigger efficiency, and correction factors, are 1--3\%. 
These are also included at the bottom of Table~\ref{tab:muon-syst}.

\begin{table}[!htb]\caption{Maximum values of the relative systematic uncertainties in the $\wmunu$ channel on the measured event yield in each $|\eta_\mu|$ interval and centrality class. Correlated uncertainties represent those that are correlated as a function of centrality or $|\eta_\mu|$. Bin--uncorrelated uncertainties represent statistical uncertainties in the background estimation, trigger efficiencies, and yield correction factors.}
\label{tab:muon-syst}
\begin{center}
\begin{tabular}{l|l}
\hline
Source & Uncertainty [\%] \\
\hline
$\mpt$ resolution & 4.0\\
QCD multi--jet background   & 2.0\\
Electroweak + $t\bar{t}$ backgrounds & 0.2\\
Muon isolation & 2.0\\
Muon reconstruction & 1.0\\
Muon \pt\ resolution & 1.0\\
Muon trigger efficiency & 0.4\\
Extrapolation correction & 0.3 \\
&\\
Total bin--correlated & 5.2\\
$\mNcoll$ determination & 9.4 \\
Total bin--uncorrelated & 3.0\\
\hline
\end{tabular}\end{center}
\end{table}

\subsection{Electron channel}

In the electron channel, the contribution due to the missing transverse momentum resolution is evaluated using the same procedure as in the muon channel. The yield variation is on average 2--5\% with a maximum deviation of 10\%. 

The uncertainty in the QCD multi--jet background
estimation arises from the choice of control region used to model the
\pt\ spectrum of fake electrons from QCD multi--jet processes. This uncertainty is
assessed by modifying the background composition of the control region in order to test the stability in the fitting
procedure under shape changes. In addition, the constraint on the
azimuthal separation between a jet -- reconstructed at the EM scale with 
$\et>25\gev$ -- and the $\mpt$ vector is loosened or tightened~\cite{Aad:2011mk}.
After applying these modifications, the altered background fractions
result in signal yield variations below 5\%.

The systematic contribution associated with the electron isolation is evaluated by varying the isolation ratio from 0.2 to 0.3. This results in an average corrected yield variation of 2\% with a maximum variation of 4\%.  

Systematic uncertainties in the electroweak background estimations are obtained from the 5\% theoretical uncertainty on each of the $\Wboson$ and $\Zboson$ boson production cross--sections. 
These uncertainties are treated as fully correlated among various $\Wboson$ and $\Zboson$ boson production processes. The resulting relative systematic uncertainty is approximately 0.2\% with the largest deviation at the level of 0.5\%.
  
The main uncertainty associated with the $C_{\Wboson^\pm}$ correction stems from possible discrepancies between data and MC simulation. In general, there are two contributions to this discrepancy: differences in the detector performance description and shortcomings in the physics model of the MC simulation that lead to distortions in the  $C_{\Wboson^\pm}$ correction given the finite binning used. 
To account for the first contribution, a result obtained in \pp\
collisions~\cite{Aad:2011mk} is used. There it was found that the electron
identification efficiencies in the data are consistent with those from
the MC simulation within a 3\% total relative uncertainty, which is applied as
a systematic uncertainty for this analysis.
The second contribution is estimated by re--weighting the signal MC sample such that the $|\eta_e|$ distribution in the simulation matches the one measured in the data. This systematic variation results in an average relative systematic uncertainty below 1\%.

The electron trigger efficiency obtained from the data using a tag--and--probe method is
compared to the efficiency from MC simulation. The efficiencies from both samples are consistent within their statistical
uncertainties. 
The statistical errors in the data are propagated
as uncertainties on the event yield, introducing a 0.2\% uncertainty.

The systematic uncertainty due to the extrapolation of the yields in the region $|\eta_e|<2.5$ is attributed to the same factors as in the muon channel (i.e. PDF uncertainties). This introduces an additional 0.2\% uncertainty in the yields from the extrapolated $|\eta_e|$ regions.  A 0.1\% uncertainty is associated with the differential production measurement in the highest $|\eta_e|$ bin.

The charge of leptons from $\wenu$ decays may be misidentified, resulting in possible misrepresentations of charge--dependent observables.  The charge misidentification probability is determined from the signal MC sample. It is below 0.2\% for $|\eta_e|<1.37$ and between 1--3\% in the highest $|\eta_e|$ region. These values are consistent with data--driven measurements~\cite{Aad:2014fxa} except in the highest $|\eta_e|$ bin, where a disagreement at the level of 50\% is found. This percentage is propagated as an uncertainty in the difference between the correction factors of each charge, resulting in a systematic uncertainty of 1.5\% and 2.0\% in the number of $\Wboson^-$ and $\Wboson^+$ boson yields, respectively, in the highest $|\eta_e|$ bin. In all other $|\eta_e|$ regions, the average relative systematic uncertainty is below $1\%$. 
The uncertainty in the charge asymmetry measurement is determined by varying the $\Wboson^-$ and $\Wboson^+$ boson yields by their respective uncertainties in opposite directions.

Table~\ref{tab:e-syst} presents a summary of the maximum values for all systematic uncertainties considered in the electron channel.  
The bin--correlated systematic uncertainties among different centrality or $|\eta_{e}|$ bins are 4.0--10.5\%. The bin--uncorrelated systematic uncertainties, which are comprised of statistical uncertainties from the background estimation, trigger efficiency, and correction factors, are 3.0--5.8\%. 
These are summarised at the bottom of Table~\ref{tab:e-syst}. 

\begin{table}[!htb]\caption{Maximum values of the relative systematic uncertainties in the $\wenu$ channel on the measured event yield in each $|\eta_e|$ interval and centrality class. Correlated uncertainties represent those that are correlated as a function of centrality or $|\eta_e|$. Uncorrelated uncertainties represent statistical uncertainties in the background estimation, trigger efficiencies, and yield correction factors.}
\label{tab:e-syst}
\begin{center}
\begin{tabular}{l|l}
\hline
Source & Uncertainty [\%] \\
\hline
$\mpt$ resolution & 10.0\\
QCD multi--jet background   & 5.0\\
Electroweak backgrounds & 0.5\\
Electron  isolation & 4.0 \\
Electron reconstruction & 3.2 \\
Electron trigger efficiency & 0.2\\
Charge misidentification & 2.0\\
Extrapolation correction & 0.2 \\
&\\
Total bin--correlated & 10.5\\
$\mNcoll$ determination & 9.4 \\
Total bin--uncorrelated & 5.8\\
\hline
\end{tabular}\end{center}
\end{table}

\subsection{Channel combination}

The results from the $\wmunu$ and $\wenu$ channels are combined in
order to increase the precision of the measurement. Although the two
channels share a common kinematic phase space, differences in their
geometrical acceptances must be considered in the
combination procedure. After verifying that the results are compatible, the two channels 
are combined using an
averaging method with weights proportional to the inverse square of the
individual uncertainties. Uncertainties treated as fully correlated between the muon and electron channels include the $\mpt$ resolution, electroweak background subtraction, and $\mNcoll$. All other sources are treated as uncorrelated.

\subsection{Theoretical predictions}
Uncertainties inherent in the PDF and EPS09 nuclear corrections are evaluated using the Hessian method to quantify the relative differences between current experimental uncertainties and central values of the PDF~\cite{Eskola:2009uj}. PDF uncertainties in the $\mathrm{Pb}$ nucleus are obtained from the weighted average of free proton and neutron PDF uncertainties. In addition, uncertainties in the renormalisation and factorisation scales are also taken into account by increasing and decreasing each scale by a factor of two and using the
maximum variation as the uncertainty in each bin. 

\section{Results}
\label{sec:results}

 The total number of background--subtracted and efficiency--corrected events in the fiducial phase space $(\pt^{\ell}>25\GeV$, $\mpt>25\GeV$, $\mt>40\GeV)$ and after extrapolation to $|\eta_{\ell}|<2.5$ is presented in Table~\ref{tab:yields} along with the ratio of $\Wboson^{+}$ and $\Wboson^{-}$ boson production.  
 
\begin{table}[!thb]\caption{Summary of the number of background--subtracted and efficiency--corrected events for $\wmunu$ and $\wenu$ events. The yields are defined in a fiducial region $\pt^{\ell}>25\GeV$, $\mpt>25\GeV$, $\mt>40\GeV$ and are extrapolated to $|\eta_{\ell}|<2.5$.}
\label{tab:yields}
\begin{center}
\resizebox{0.39\textwidth}{!}{
\begin{tabular}{ll}
\hline
\pho{} & $\wmunu$ \\
\hline
$\Wboson^{+}$ &             $5870\phantom{.}~\pm100\phantom{.}~\stat~\pm90\phantom{0}\phantom{.}~\syst$ \\
$\Wboson^{-}$ &             $5680\phantom{.}~\pm100\phantom{.}~\stat~\pm80\phantom{0}\phantom{.}~\syst$     \\
$\Wboson^{+}/\Wboson^{-}$ & $1.03\phantom{0}~\pm0.03~\stat~\pm0.02~\syst$ \\
\hline
\pho{} & $\wenu$ \\
\hline
 $\Wboson^{+}$ &             $5760\phantom{.}~\pm150\phantom{.}~\stat~\pm90\phantom{0}\phantom{.}~\syst$ \\
 $\Wboson^{-}$ &             $5650\phantom{.}~\pm150\phantom{.}~\stat~\pm110\phantom{.}~\syst$ \\   
 $\Wboson^{+}/\Wboson^{-}$ & $1.02\phantom{0}~\pm0.04~\stat~\pm0.01~\syst$ \\
\hline
\end{tabular}}
\end{center}
\end{table}

\begin{figure}[!thb]
\begin{center}

\resizebox{0.48\textwidth}{!}{
    \includegraphics[]{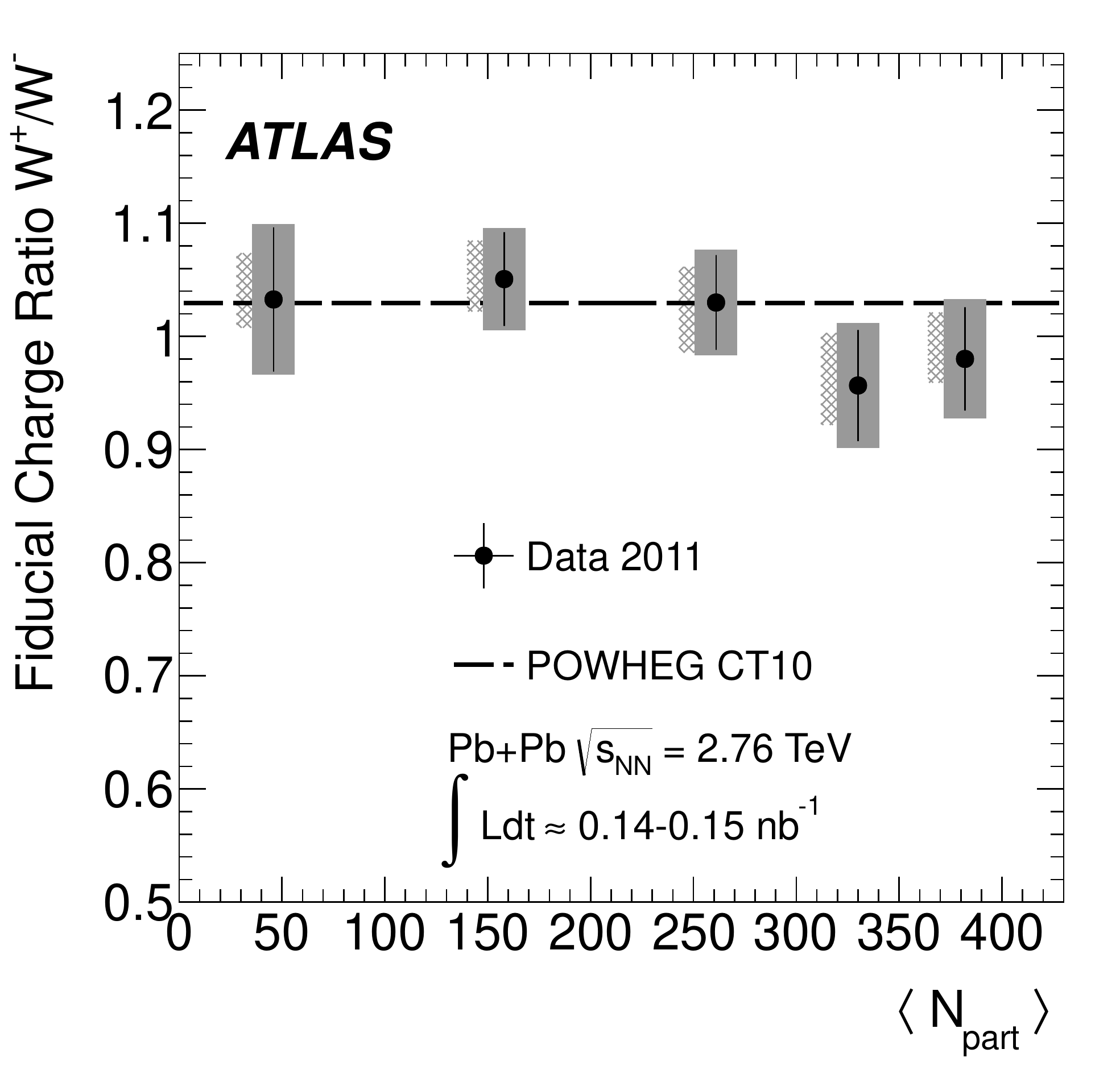}
}

\end{center}
\caption{ Ratio of $\Wboson^{+}$ and $\Wboson^{-}$ candidates (from $\wlnu$)  as a function of $\mNpart$.  
  The kinematic requirements are $\pt^{\ell}>25\gev$, $\mpt>25\gev$, $\mt>40\gev$, and $|\eta_{\ell}|<2.5$. Also shown is a QCD NLO prediction from \powheg. Statistical uncertainties are shown as black bars. The filled grey boxes represent statistical and bin--uncorrelated systematic uncertainties added in quadrature, whereas the grey--hatched boxes represent bin--correlated uncertainties and are offset for clarity. }
\label{fig:ratio}       
\end{figure}

\noindent The corrected yields from each channel are consistent.  
Moreover, the contributions from \nn\ and \pn\ collisions are evident.
\pp\ collisions alone would result in a ratio of $\Wboson^{+}$ and $\Wboson^{-}$ bosons significantly above unity, but in $\PbPb$ collisions, the 
larger number of $d$ valence quarks in the neutron increases $\Wboson^{-}$ production, driving the ratio closer to one. 
This is supported by Fig.~\ref{fig:ratio}, which presents the fiducial charge ratio as a function of $\mNpart$ for the combined muon and electron channels.  

\begin{figure}[!thb]
\begin{center}
\includegraphics[width=0.48\textwidth]{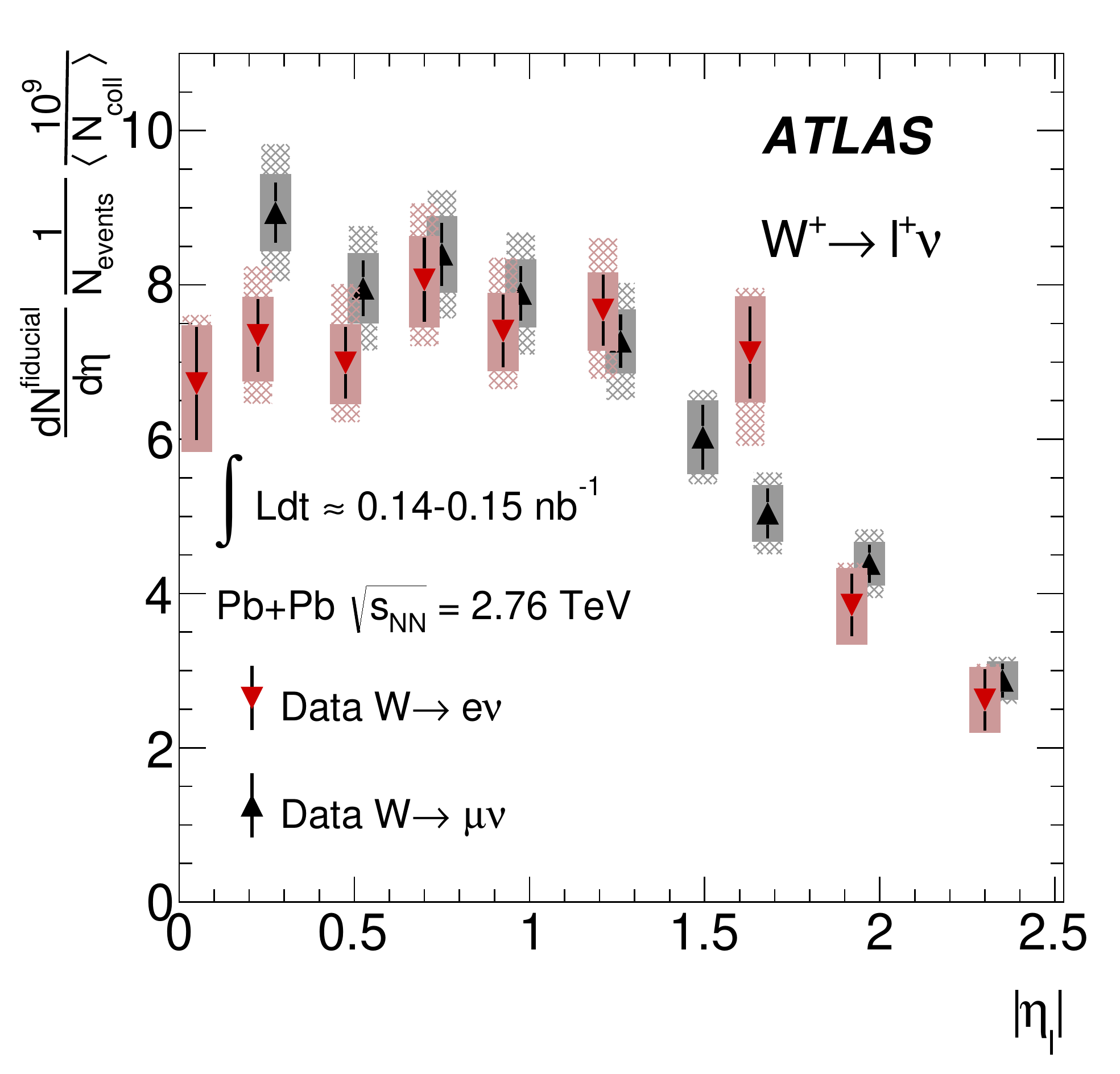}
\includegraphics[width=0.48\textwidth]{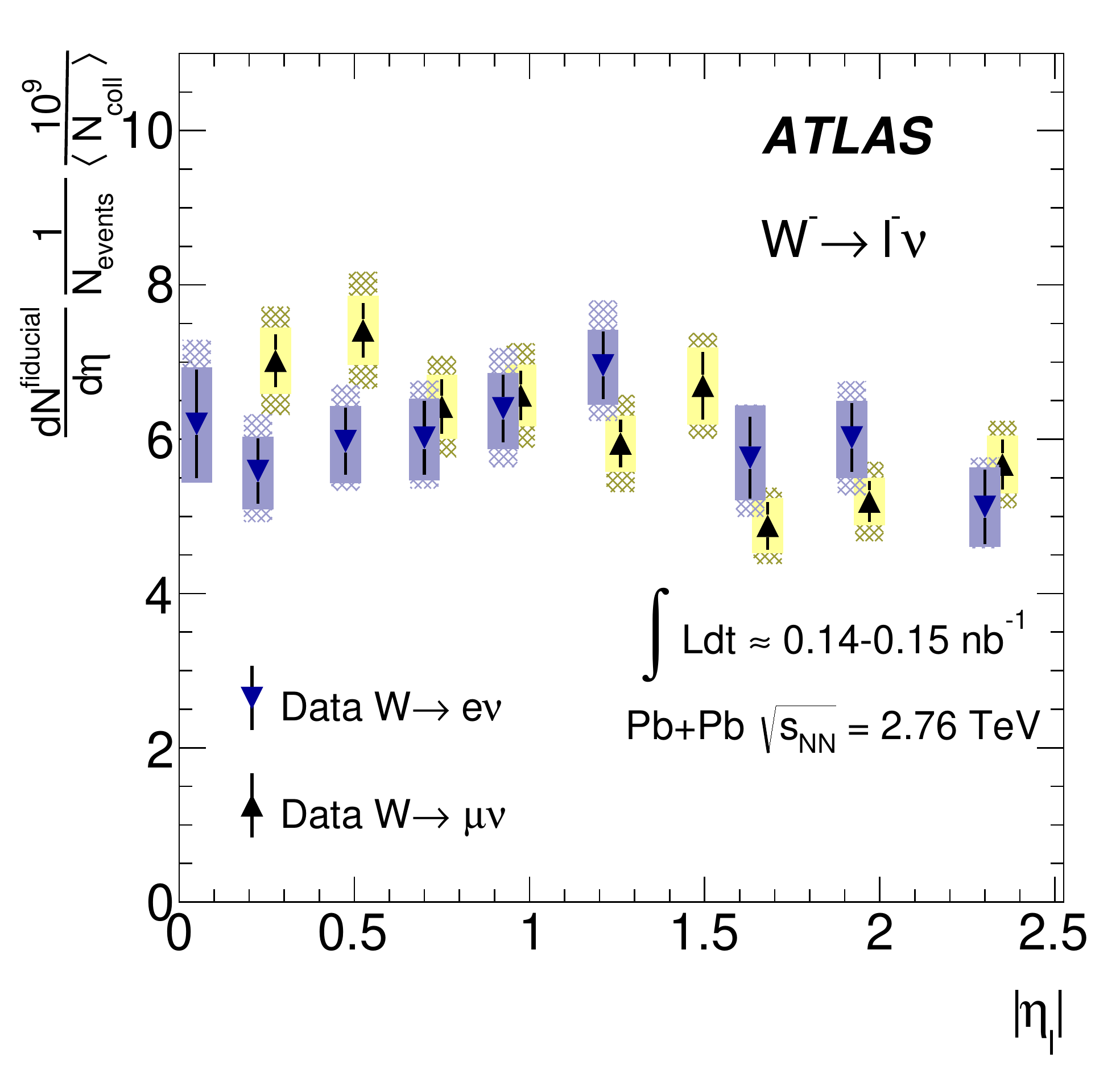}
\caption{Differential production yields per binary collision for $\Wboson^+$~(top) and $\Wboson^-$~(bottom) events from electron and muon channels. Due to acceptance the first bin in the muon channel and the seventh bin in the electron channel are not covered. Muon
 points are shifted horizontally for visibility. The kinematic requirements are $\pt^{\ell}>25\gev$, $\mpt>25\gev$, and $\mt>40\gev$. Statistical 
errors are shown as black bars, whereas bin--uncorrelated systematic and statistical uncertainties added in quadrature are shown as the filled error box. Bin--correlated uncertainties are shown as the hatched boxes. These include uncertainties from $\mNcoll$.
}\label{fig:e-mu-comp}
\end{center}
\end{figure}

Figure~\ref{fig:e-mu-comp} shows a comparison between the differential production yields per binary collision for the muon and electron channels, separately, as a function of $|\eta_\ell|$ for $\Wboson^+$ and $\Wboson^-$. A good agreement is found between the two decay modes.
In both decay channels, the distribution from $\Wboson^+$ bosons steeply falls at large $|\eta_\ell|$, whereas this is not the case for $\Wboson^-$ events. This behaviour is understood 
and is further discussed below in connection to the charge asymmetry. 

\begin{figure}[!thb]
\begin{center}

\resizebox{0.48\textwidth}{!}{
    \includegraphics[]{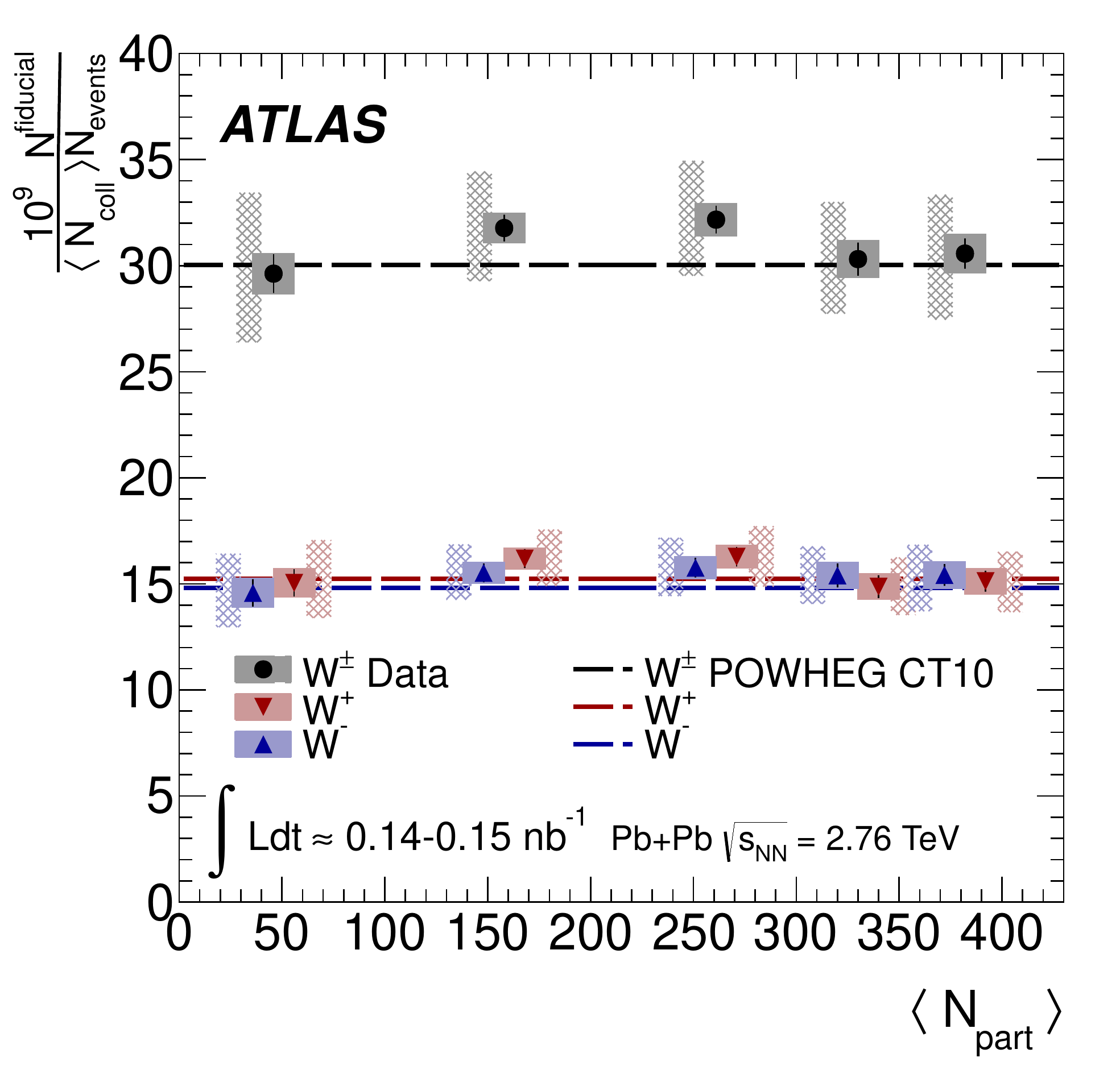}

}
\end{center}

\caption{ $\Wboson$ boson production yield per binary collision as a function of the mean number of participants $\mNpart $
for $\Wboson^{+}$, $\Wboson^{-}$, and $\Wboson^{\pm}$ bosons for combined muon and electron channels. The kinematic requirements are $\pt^{\ell}>25\gev$, $\mpt>25\gev$, $\mt>40\gev$, and $|\eta_{\ell}|<2.5$.
Statistical errors are shown as black bars, whereas bin--uncorrelated systematic and statistical uncertainties added in quadrature are shown as the filled error box. Bin--correlated uncertainties are shown as the hatched boxes and are offset for clarity. These include uncertainties from $\mNcoll$. Also shown is an NLO QCD prediction.  
}
\label{fig:w-yield}       
\end{figure}

Figure~\ref{fig:w-yield} presents the $\Wboson$
boson production yield per binary collision for each charge separately as
well as inclusively as a function of $\mNpart$ for the combined data.
Also shown are comparisons to QCD NLO predictions.  
The NLO predictions are consistent with the data for both the charge ratio, as shown in Fig.~\ref{fig:ratio}, and production yields in Fig.~\ref{fig:w-yield}.

\begin{figure}[!thb]
\begin{center}
\resizebox{0.48\textwidth}{!}{
    \includegraphics[]{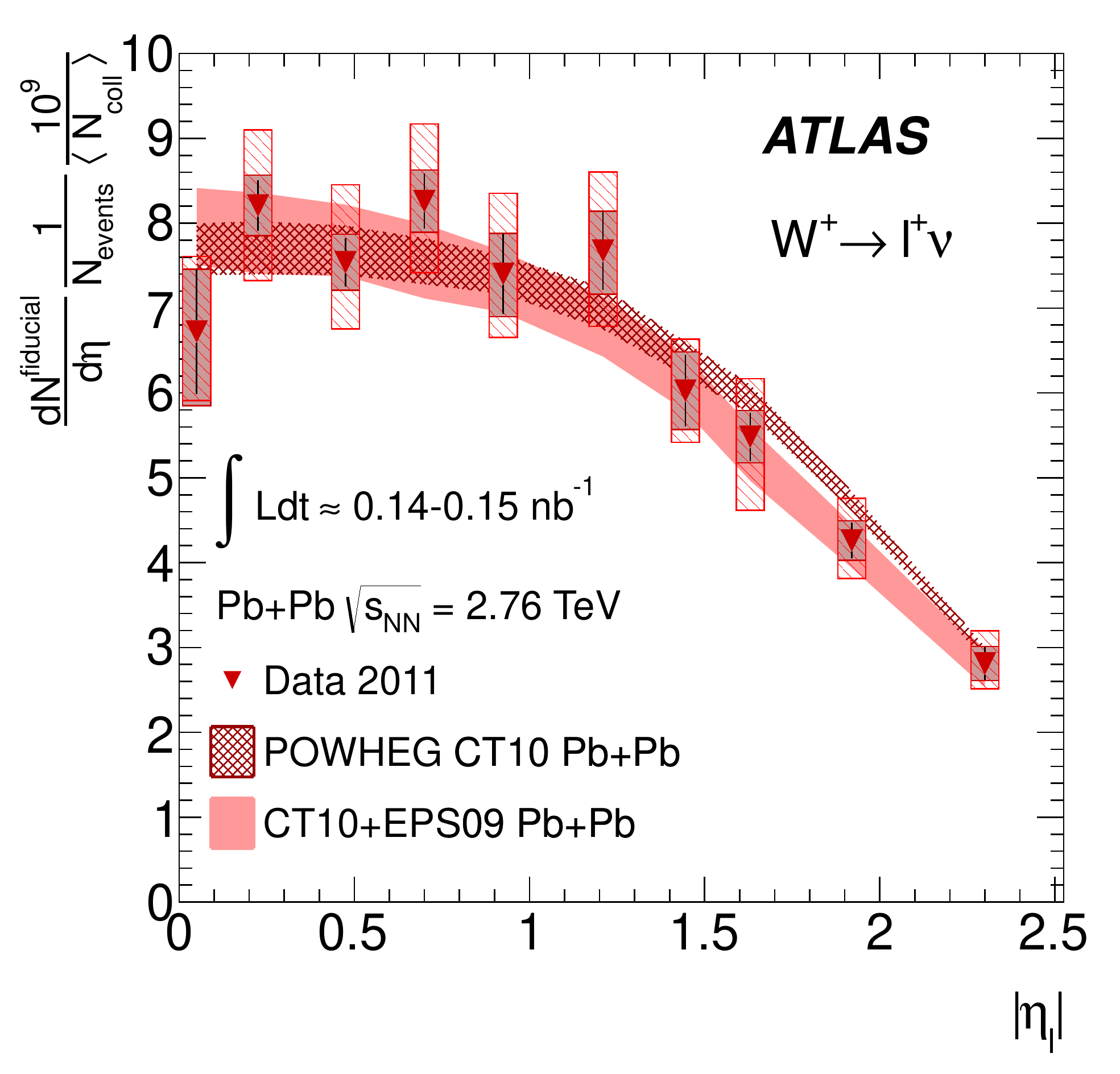}
}
\resizebox{0.48\textwidth}{!}{
    \includegraphics[]{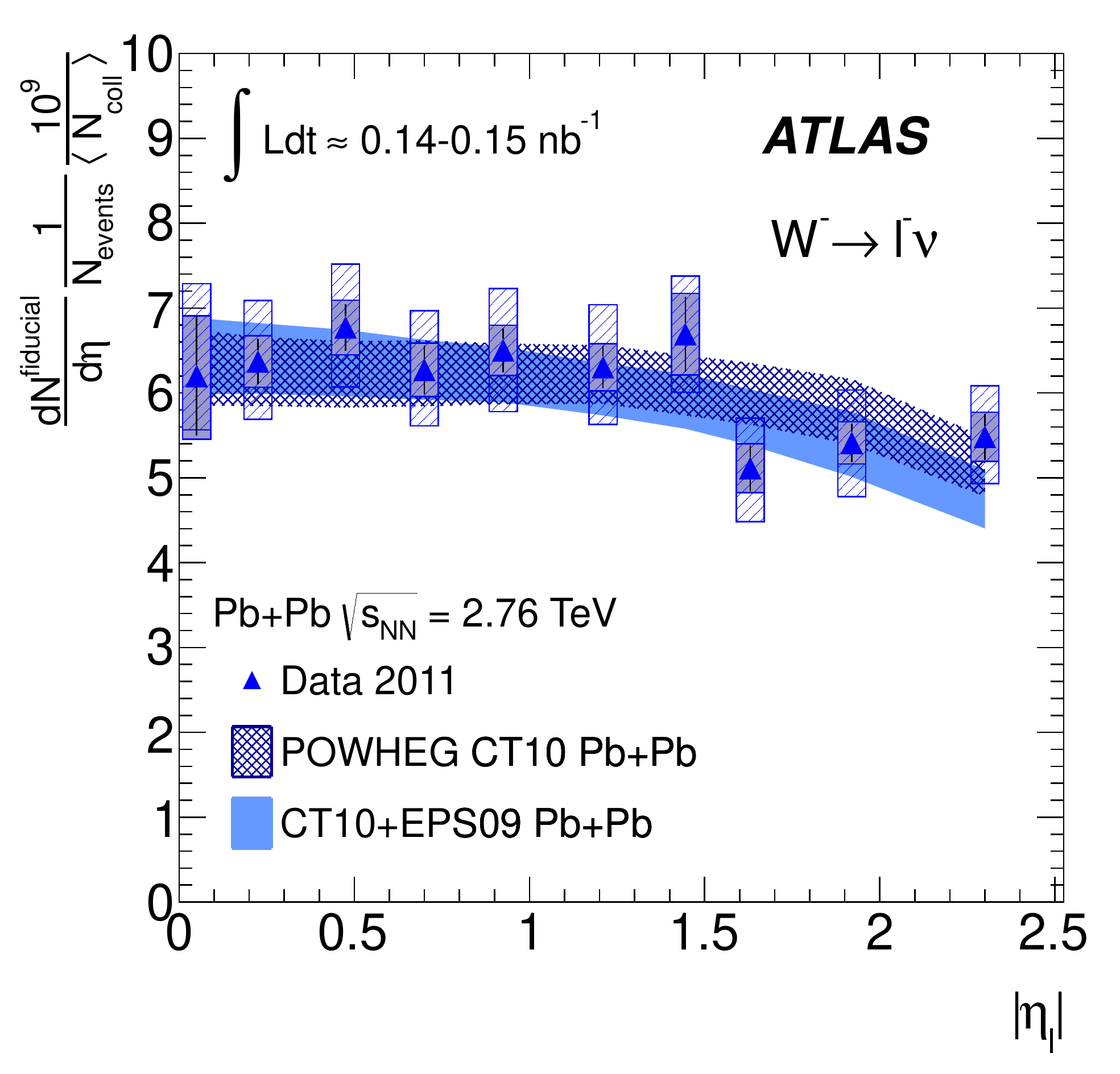}
}
\end{center}
\caption{ Differential production yield per binary collision for $\Wboson^{+}$~(top) and $\Wboson^{-}$~(bottom) events integrated 
over all centralities and compared to NLO QCD theoretical predictions with~(CT10+EPS09) and without~(CT10) nuclear corrections. 
The kinematic requirements are $\pt^{\ell}>25\gev$, $\mpt>25\gev$, and $\mt>40\gev$. Statistical errors are shown as black bars, whereas bin--uncorrelated systematic and statistical uncertainties added in quadrature are shown as the filled error box. Bin--correlated uncertainties are shown as the hatched boxes. These include uncertainties from $\mNcoll$.
The PDF uncertainties in both the CT10+EPS09 and CT10 predictions are derived from the PDF error eigensets. The total theoretical uncertainty also includes uncertainties in the renormalisation and factorisation scales used in the cross-section calculations.
 }
\label{fig:yield-eta}      
\end{figure}

As with other heavy--ion electroweak boson measurements, $\Wboson$ boson production yields per binary nucleon--nucleon collision are independent of centrality.  
This suggests that the $\Wboson$ boson can be used for benchmarking energy--loss processes in a QGP.
Thus, when produced in association with jets, $\Wboson$ boson production introduces an additional avenue for exploring in--medium modifications -- energy loss due to multiple scattering and gluon radiation  -- to 
energetic partons traversing the heavy--ion medium.

\begin{figure}[!thbp]
\begin{center}
\resizebox{0.48\textwidth}{!}{
    \includegraphics[]{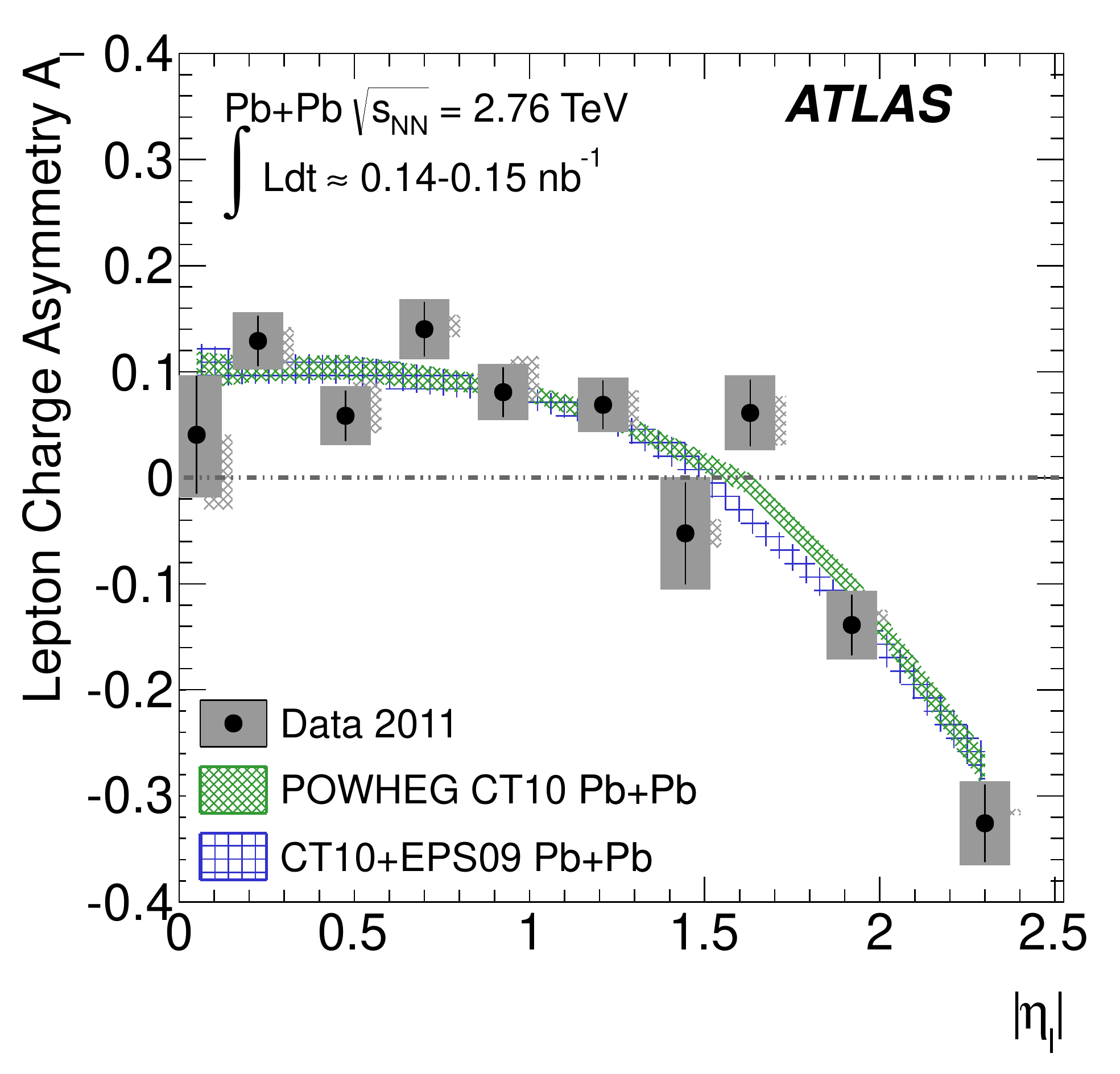}
}
\end{center}
\caption{ The lepton charge asymmetry $A_{\ell}$ from $\Wpm$ bosons as a function of absolute pseudorapidity compared to theoretical predictions from the CT10 and CT10+EPS09 NLO PDF sets. 
The kinematic requirements are $\pt^{\ell}>25\gev$, $\mpt>25\gev$, and $\mt>40\gev$. Statistical uncertainties are shown as black bars, whereas bin--uncorrelated systematic and statistical uncertainties added in quadrature are shown as the filled error box. Correlated scaling uncertainties are shown as the hatched boxes and are offset for clarity.
The PDF uncertainties in both the CT10+EPS09 and CT10 predictions are derived from the PDF error eigensets. The total theoretical uncertainty also includes uncertainties in the renormalisation and factorisation scales used in the cross-section calculations.
}
\label{fig:asymmetry}       
\end{figure}

Nuclear modifications to the PDF are explored in 
Figs.~\ref{fig:yield-eta} and~\ref{fig:asymmetry}, which present the differential $\wlnu$ production yields per binary nucleon--nucleon collision and the lepton charge asymmetry, respectively, as a function of $|\eta_{\ell}|$. Each figure includes NLO predictions with the CT10 PDF set, both with and without EPS09 nuclear corrections.
The EPS09 corrections incorporate modifications to the PDF that account for 
contributions from shadowing, anti--shadowing, the EMC--effect, and Fermi--motion~\cite{Paukkunen:2010qg}.  

Both the CT10 and CT10+EPS09 predictions in Figs.~\ref{fig:yield-eta} and \ref{fig:asymmetry} describe the 
data well. Therefore, at the current level of theoretical and experimental precision, this measurement is insensitive to    
nuclear modifications to the PDF. 
Fig.~\ref{fig:asymmetry} also exhibits a sign--change of the charge asymmetry at $|\eta_{\ell}|\approx1.5$, behaviour hitherto only observed at $|\eta_{\ell}|>3$ in $\pp$ measurements at $7\tev$~\cite{Aad:2011dm,Aaij:2012vn}. 
The negative asymmetry is attributable to the $V - A$ structure of $\Wboson$ boson decays, in which the decay angle of the charged lepton is anisotropic and a larger fraction of 
negatively charged leptons are produced at forward $|\eta_{\ell}|$. The larger fraction of $\wlnum$ events in $\PbPb$ compared to $\pp$ collisions results in 
a sign--change of the asymmetry that can be observed within the $|\eta_{\ell}|$ acceptance of the measurement.
This behaviour is in accordance with the NLO QCD predictions. 

\section{Summary and conclusions}
\label{sec:summary}
The measurements of $\Wboson^{\pm}$ boson production in \PbPb\ collisions at \energy\ are presented using data corresponding to 
an integrated luminosity of $0.14-0.15\;\inb$ collected with the ATLAS detector at the LHC. The $\Wboson^{\pm}$ boson candidates 
are selected using muons or electrons in the final state in the fiducial region defined by  $\pt^{\ell}>25\GeV$, $\mpt>25\GeV$, $\mt>40\GeV$ and $0.1<|\eta_{\mu}|<2.4$ for muons and $|\eta_e|<2.47$, excluding the transition region, for electrons. 
After background subtraction, correction, and extrapolation to a pseudorapidity coverage of $|\eta_\ell|<2.5$, the numbers of events reported in 
each channel are consistent.

The $\Wboson$ boson production yields are presented as a function of $\mNpart$ and $|\eta_\ell|$. 
These yields, scaled by $1/\langle \Ncoll\rangle$, are independent of centrality and in agreement with NLO QCD predictions.
The lepton charge asymmetry from $\Wboson^\pm$ boson decays
differs from measurements in \pp\ collisions. This
is expected since in \PbPb\ collisions there is an additional neutron component contributing 
to $\Wboson$ boson production.
The lepton charge asymmetry agrees well with theoretical predictions using QCD at NLO with CT10 PDF sets with and without EPS09 nuclear corrections. 
The nuclear corrections account for modifications that are not present in the PDF of free nucleons.
However, further improvements in the experimental precision and uncertainties in the theory are needed to establish the existence of nuclear effects.
The results presented here clearly indicate that in events associated with a jet,
$\Wboson$ bosons are an excellent tool for evaluating jet energy--loss in a QGP.
Moreover, it was demonstrated that $\Wboson$ bosons can be used to study PDFs in multi--nucleon systems. 
With improved statistical and systematic precision, along with additional data from different colliding systems such as \pPb, it will be possible
to decisively evaluate the extent of nuclear effects on PDFs and to further test theoretical predictions. 

\section*{Acknowledgements}

We thank CERN for the very successful operation of the LHC, as well as the
support staff from our institutions without whom ATLAS could not be
operated efficiently.

We acknowledge the support of ANPCyT, Argentina; YerPhI, Armenia; ARC,
Australia; BMWF and FWF, Austria; ANAS, Azerbaijan; SSTC, Belarus; CNPq and FAPESP,
Brazil; NSERC, NRC and CFI, Canada; CERN; CONICYT, Chile; CAS, MOST and NSFC,
China; COLCIENCIAS, Colombia; MSMT CR, MPO CR and VSC CR, Czech Republic;
DNRF, DNSRC and Lundbeck Foundation, Denmark; EPLANET, ERC and NSRF, European Union;
IN2P3-CNRS, CEA-DSM/IRFU, France; GNSF, Georgia; BMBF, DFG, HGF, MPG and AvH
Foundation, Germany; GSRT and NSRF, Greece; ISF, MINERVA, GIF, I-CORE and Benoziyo Center,
Israel; INFN, Italy; MEXT and JSPS, Japan; CNRST, Morocco; FOM and NWO,
Netherlands; BRF and RCN, Norway; MNiSW and NCN, Poland; GRICES and FCT, Portugal; MNE/IFA, Romania; MES of Russia and ROSATOM, Russian Federation; JINR; MSTD,
Serbia; MSSR, Slovakia; ARRS and MIZ\v{S}, Slovenia; DST/NRF, South Africa;
MINECO, Spain; SRC and Wallenberg Foundation, Sweden; SER, SNSF and Cantons of
Bern and Geneva, Switzerland; NSC, Taiwan; TAEK, Turkey; STFC, the Royal
Society and Leverhulme Trust, United Kingdom; DOE and NSF, United States of
America.

The crucial computing support from all WLCG partners is acknowledged
gratefully, in particular from CERN and the ATLAS Tier-1 facilities at
TRIUMF (Canada), NDGF (Denmark, Norway, Sweden), CC-IN2P3 (France),
KIT/GridKA (Germany), INFN-CNAF (Italy), NL-T1 (Netherlands), PIC (Spain),
ASGC (Taiwan), RAL (UK) and BNL (USA) and in the Tier-2 facilities
worldwide.

\FloatBarrier

\flushend

\bibliographystyle{modspphys}
\raggedright
\bibliography{wlnu_paper}
\justifying
\onecolumn
\clearpage
\begin{flushleft}
{\Large The ATLAS Collaboration}

\bigskip

G.~Aad$^{\rm 84}$,
B.~Abbott$^{\rm 112}$,
J.~Abdallah$^{\rm 152}$,
S.~Abdel~Khalek$^{\rm 116}$,
O.~Abdinov$^{\rm 11}$,
R.~Aben$^{\rm 106}$,
B.~Abi$^{\rm 113}$,
M.~Abolins$^{\rm 89}$,
O.S.~AbouZeid$^{\rm 159}$,
H.~Abramowicz$^{\rm 154}$
% ...
\bigskip
\\
% ...
$^{11}$ Institute of Physics, Azerbaijan Academy of Sciences, Baku, Azerbaijan\\
% ...
$^{84}$ CPPM, Aix-Marseille Universit{\'e} and CNRS/IN2P3, Marseille, France\\
% ...
$^{89}$ Department of Physics and Astronomy, Michigan State University, East Lansing MI, United States of America\\
% ...
$^{106}$ Nikhef National Institute for Subatomic Physics and University of Amsterdam, Amsterdam, Netherlands\\
% ...
$^{112}$ Homer L. Dodge Department of Physics and Astronomy, University of Oklahoma, Norman OK, United States of America\\
$^{113}$ Department of Physics, Oklahoma State University, Stillwater OK, United States of America\\
% ...
$^{116}$ LAL, Universit{\'e} Paris-Sud and CNRS/IN2P3, Orsay, France\\
% ...
$^{152}$ Institute of Physics, Academia Sinica, Taipei, Taiwan\\
% ...
$^{154}$ Raymond and Beverly Sackler School of Physics and Astronomy, Tel Aviv University, Tel Aviv, Israel\\
% ...
$^{159}$ Department of Physics, University of Toronto, Toronto ON, Canada\\
% ...
\end{flushleft}

\end{document}